\documentclass[conference]{IEEEtran}
\IEEEoverridecommandlockouts
\usepackage{fancyhdr}

\usepackage{cite}
\usepackage{amsmath,amssymb,amsfonts}
\usepackage{algorithmic}
\usepackage{graphicx}
\usepackage{textcomp}
\usepackage{xcolor}
\usepackage{booktabs}
\usepackage{subcaption}
\usepackage{multirow}
\usepackage{siunitx}
\usepackage{hyperref}
\usepackage[capitalise]{cleveref}

\graphicspath{{imgs/}}

\newcommand{\HaccParticlesM}{281}  
\newcommand{\HaccRawGB}{3.4}  
\newcommand{\FireTwoParticlesM}{268}  
\newcommand{\FireTwoRawGB}{3.2}  
\newcommand{\HaccDomain}{256}  
\newcommand{\EvalFrames}{80}
\newcommand{\OrbitFar}{1.0}
\newcommand{\OrbitMid}{0.7}
\newcommand{\OrbitNear}{0.5}
\newcommand{\EvalTotalFrames}{240}  

\newcommand{\SbPSNR}{28.80}  
\newcommand{\SbSizeMB}{11.1}  
\newcommand{\SbGaussians}{171k}  

\newcommand{\SbCR}{290}  
\newcommand{\SbSSIM}{0.7723}  

\newcommand{\EbPSNR}{30.03}  
\newcommand{\EbSizeMB}{49.8}  
\newcommand{\EbGaussians}{769k}  

\newcommand{\EbCR}{65}  

\newcommand{\AblNoVmPSNR}{23.30}  
\newcommand{\AblNoVmDelta}{-5.51}  
\newcommand{\AblNoVmSizeMB}{14.4}
\newcommand{\AblNoVmGaussians}{223k}  
\newcommand{\AblNoVmGaussianPct}{30}  
\newcommand{\AblNoProgPSNR}{26.42}
\newcommand{\AblNoProgDelta}{-2.39}
\newcommand{\AblNoProgSizeMB}{11.0}
\newcommand{\AblNoProgGaussians}{169k}
\newcommand{\AblNoOrbitPSNR}{28.04}
\newcommand{\AblNoOrbitDelta}{-0.77}
\newcommand{\AblNoOrbitSizeMB}{12.4}
\newcommand{\AblNoOrbitGaussians}{191k}

\newcommand{\AblFtNoDensifyPSNR}{28.61}
\newcommand{\AblFtNoDensifyDelta}{-1.42}
\newcommand{\AblFtNoDensifySizeMB}{57.9}
\newcommand{\AblFtPureDssimPSNR}{26.60}
\newcommand{\AblFtPureDssimDelta}{-3.43}
\newcommand{\AblFtPureDssimSizeMB}{298.2}
\newcommand{\AblFtPureDssimBloat}{6}  
\newcommand{\AblFtDssimBlendPSNR}{28.26}
\newcommand{\AblFtDssimBlendDelta}{-1.77}
\newcommand{\AblFtDssimBlendSizeMB}{54.2}

\newcommand{\BlkTwoMergedPSNR}{28.99}
\newcommand{\BlkTwoFtPSNR}{29.72}
\newcommand{\BlkTwoDelta}{+0.92}
\newcommand{\BlkTwoGaussians}{543k}  
\newcommand{\BlkTwoSizeMB}{35.2}
\newcommand{\BlkTwoSSIM}{0.8076}  
\newcommand{\BlkFourMergedPSNR}{28.85}
\newcommand{\BlkFourFtPSNR}{29.94}
\newcommand{\BlkFourDelta}{+1.14}
\newcommand{\BlkFourGaussians}{601k}  
\newcommand{\BlkFourSizeMB}{39.0}
\newcommand{\BlkFourSSIM}{0.8195}  
\newcommand{\BlkEightMergedPSNR}{28.32}
\newcommand{\BlkEightFtPSNR}{30.03}
\newcommand{\BlkEightDelta}{+1.23}
\newcommand{\BlkEightGaussians}{769k}  
\newcommand{\BlkEightSizeMB}{49.8}
\newcommand{\BlkEightSSIM}{0.8241}  
\newcommand{\BlkSixteenMergedPSNR}{26.80}
\newcommand{\BlkSixteenFtPSNR}{30.08}
\newcommand{\BlkSixteenDelta}{+1.28}
\newcommand{\BlkSixteenGaussians}{1{,}037k}  
\newcommand{\BlkSixteenSizeMB}{67.2}
\newcommand{\BlkSixteenSSIM}{0.8271}  
\newcommand{\BlockGainMax}{1.28}

\newcommand{\BlkDistOneFar}{26.32}

\newcommand{\BlkDistOneNear}{26.05}

\newcommand{\BlkDistEightFar}{27.35}

\newcommand{\BlkDistEightNear}{27.54}

\newcommand{\BlkDistNearGain}{+1.49}
\newcommand{\BlkDistFarGain}{+1.04}

\newcommand{\ScaleOneGaussians}{171k}

\newcommand{\ScaleSixteenParticlesM}{17.6}
\newcommand{\ScaleSixteenGaussians}{94k}

\newcommand{\SzPsnrAtSbCR}{21.2}  
\newcommand{\SzPsnrAtEbCR}{24.7}  
\newcommand{\LcpPsnrAtSbCR}{17.9}  
\newcommand{\LcpPsnrAtEbCR}{20.3}  
\newcommand{\GainSzSb}{+7.6}  
\newcommand{\GainSzEb}{+5.4}  
\newcommand{\GainLcpSb}{+10.9}  
\newcommand{\GainLcpEb}{+9.7}  
\newcommand{\GainSzLow}{5}  
\newcommand{\GainSzHigh}{8}  
\newcommand{\SzCrAtEbEight}{15}  
\newcommand{\LcpCrAtEbEight}{45}  
\newcommand{\ArtifactSzCR}{63}  
\newcommand{\SzSizeAtSbCR}{11}  
\newcommand{\SzSizeAtEbCR}{50}  
\newcommand{\QualSzPSNR}{24.73}  

\newcommand{\GsFpsOneK}{662}  
\newcommand{\GsFpsFourK}{518}  
\newcommand{\PvFps}{0.28}  
\newcommand{\PvSecPerFrame}{3.6}
\newcommand{\PerfPvVizMs}{3{,}474}  
\newcommand{\PerfPvCamMs}{129}  
\newcommand{\PerfPvCombMs}{3{,}603}  
\newcommand{\PerfGsVmMs}{1.21}  
\newcommand{\PerfGsRastMs}{0.38}  
\newcommand{\PerfGsCombMs}{1.51}  
\newcommand{\PerfPvVizFourKMs}{3{,}518}  
\newcommand{\PerfPvCamFourKMs}{129}  
\newcommand{\PerfPvCombFourKMs}{3{,}648}  
\newcommand{\PerfGsVmFourKMs}{1.20}
\newcommand{\PerfGsRastFourKMs}{0.77}
\newcommand{\PerfGsCombFourKMs}{1.93}
\newcommand{\SpeedupOneK}{2{,}386}  
\newcommand{\SpeedupFourK}{1{,}890}  
\newcommand{\SpeedupRounded}{2{,}300}  
\newcommand{\SpeedupVizOneK}{2{,}871}
\newcommand{\SpeedupCamOneK}{339}
\newcommand{\SpeedupVizFourK}{2{,}932}
\newcommand{\SpeedupCamFourK}{168}
\newcommand{\InferMemMB}{462.2}
\newcommand{\TrainPeakGB}{8.5}
\newcommand{\TrainTimeMin}{67}  
\newcommand{\FtTimeMin}{15}  
\newcommand{\PvLoadSec}{39.4}
\newcommand{\GsLoadMs}{41}
\newcommand{\LoadSpeedup}{968}  
\newcommand{\MergeTimeSec}{0.11}  

\newcommand{\HaccGenZeroPSNR}{27.33}
\newcommand{\HaccGenZeroSizeMB}{11.7}
\newcommand{\HaccGenZeroGaussians}{180k}  
\newcommand{\HaccGenZeroCR}{264}  
\newcommand{\HaccGenOnePSNR}{27.37}
\newcommand{\HaccGenOneSizeMB}{11.4}
\newcommand{\HaccGenOneGaussians}{175k}  
\newcommand{\HaccGenOneCR}{270}  
\newcommand{\HaccGenTwoPSNR}{27.33}
\newcommand{\HaccGenTwoSizeMB}{10.8}
\newcommand{\HaccGenTwoGaussians}{166k}  
\newcommand{\HaccGenTwoCR}{286}  
\newcommand{\HaccGenThreePSNR}{27.42}
\newcommand{\HaccGenThreeSizeMB}{11.3}
\newcommand{\HaccGenThreeGaussians}{174k}  
\newcommand{\HaccGenThreeCR}{272}  
\newcommand{\HaccGenMeanPSNR}{27.36}
\newcommand{\HaccGenStdPSNR}{0.04}
\newcommand{\HaccGenCRLow}{264}
\newcommand{\HaccGenCRHigh}{286}
\newcommand{\HaccGenSizeLow}{10.8}
\newcommand{\HaccGenSizeHigh}{11.7}
\newcommand{\HaccGenMeanSizeMB}{11.3}
\newcommand{\HaccGenMeanGaussians}{174k}
\newcommand{\HaccGenMeanCR}{273}
\newcommand{\HaccGenParticlesM}{268}
\newcommand{\FireTwoPSNR}{29.27}
\newcommand{\FireTwoSizeMB}{5.3}
\newcommand{\FireTwoGaussians}{82k}  
\newcommand{\FireTwoCR}{577}  
\newcommand{\FireTwoDomain}{116{,}960}  
\newcommand{\GenPsnrLow}{27}  
\newcommand{\GenPsnrHigh}{29}  
\newcommand{\GenCRLow}{264}
\newcommand{\GenCRHigh}{577}

\newcommand{\RecVZeroCorr}{0.928}
\newcommand{\RecVZeroMAE}{0.718}
\newcommand{\RecVZeroNNMean}{0.250}
\newcommand{\RecVOneCorr}{0.437}
\newcommand{\RecVOneMAE}{1.172}
\newcommand{\RecVOneNNMean}{1.216}
\newcommand{\RecVFourCorr}{0.917}
\newcommand{\RecVFourMAE}{0.701}
\newcommand{\RecVFourNNMean}{0.281}
\newcommand{\RecVSixCorr}{0.003}
\newcommand{\RecVSixMAE}{1.924}
\newcommand{\RecVSixNNMean}{0.061}
\newcommand{\RecGtNNMean}{0.665}
\newcommand{\RecVZeroXiDev}{8.6}
\newcommand{\RecVOneXiDev}{4.0}
\newcommand{\RecVFourXiDev}{8.9}
\newcommand{\RecVSixXiDev}{154.1}

\newcommand{\RecBlkOneCorr}{0.897}
\newcommand{\RecBlkOneMAE}{0.906}
\newcommand{\RecBlkOneNNMean}{0.161}
\newcommand{\RecBlkOneXiDev}{15.3}  
\newcommand{\RecBlkTwoCorr}{0.916}
\newcommand{\RecBlkTwoMAE}{0.815}
\newcommand{\RecBlkTwoNNMean}{0.186}
\newcommand{\RecBlkTwoXiDev}{9.3}  
\newcommand{\RecBlkFourCorr}{0.923}
\newcommand{\RecBlkFourMAE}{0.760}
\newcommand{\RecBlkFourNNMean}{0.218}
\newcommand{\RecBlkFourXiDev}{8.3}  
\newcommand{\RecBlkEightCorr}{0.928}
\newcommand{\RecBlkEightMAE}{0.718}
\newcommand{\RecBlkEightNNMean}{0.250}
\newcommand{\RecBlkEightXiDev}{8.6}  
\newcommand{\RecBlkSixteenCorr}{0.928}
\newcommand{\RecBlkSixteenMAE}{0.681}
\newcommand{\RecBlkSixteenNNMean}{0.292}
\newcommand{\RecBlkSixteenXiDev}{7.8}  
\newcommand{\RecNNPctOfGT}{44}  
\newcommand{\RecCorrThreshold}{0.92}  
\newcommand{\RecXiTenPctR}{10}  
\newcommand{\RecPkTenPctK}{2.63}  
\newcommand{\RecTimeSec}{74.5}  

\newcommand{\ThreeWayGsAvg}{27.56}
\newcommand{\ThreeWayRecAvg}{16.95}
\newcommand{\ThreeWayGap}{11}  

\newcommand{\DssimSizeIncreasePct}{68}  
\newcommand{\ShSavingsPct}{59}  
\newcommand{\BlockSizeIncFourToEightPct}{28}  
\newcommand{\BlockSatDelta}{+0.09}  

\newcommand{\VmParams}{4{,}610}  
\newcommand{\VmParamsApprox}{4{,}600}  
\newcommand{\VmArch}{4 \to 64 \to 64 \to 2}
\newcommand{\InitPoints}{200{,}000}  
\newcommand{\DefaultR}{0.01}  
\newcommand{\DefaultAlpha}{0.05}  
\newcommand{\BetaConc}{3.0}
\newcommand{\RadiusMin}{0.0025}
\newcommand{\RadiusMax}{0.0175}
\newcommand{\OpacityMin}{0.0125}
\newcommand{\OpacityMax}{0.0875}
\newcommand{\DeltaSMax}{0.1}  
\newcommand{\DeltaOMax}{0.3}  
\newcommand{\InitEpsilon}{0.4}  
\newcommand{\FtDeltaSMax}{0.3}  
\newcommand{\FtDeltaOMax}{0.8}  
\newcommand{\FtEpsilon}{0.1}  
\newcommand{\FtIters}{60{,}000}  
\newcommand{\FtDensifyUntil}{30{,}000}  
\newcommand{\StageOneRes}{1920 {\times} 1080}  
\newcommand{\StageOneIters}{12{,}000}  
\newcommand{\StageTwoRes}{5760 {\times} 3240}  
\newcommand{\StageTwoIters}{27{,}000}  
\newcommand{\StageTwoExtIters}{15{,}000}  
\newcommand{\StageTwoMixIters}{12{,}000}  
\newcommand{\InternalMixPct}{20}  
\newcommand{\TotalIters}{39{,}000}  
\newcommand{\EvalRes}{1920 {\times} 1080}  
\newcommand{\RecGridSize}{256}  

\newcommand{\GenModelGaussians}{171{,}346}
\newcommand{\GenReproPSNR}{28.82}      
\newcommand{\GenOrbitFarPSNR}{26.20}      
\newcommand{\GenOrbitTrainFarPSNR}{26.32} 
\newcommand{\GenOrbitInterpHiPSNR}{26.33} 
\newcommand{\GenOrbitTrainMidPSNR}{26.43} 
\newcommand{\GenOrbitInterpLoPSNR}{26.37} 
\newcommand{\GenOrbitTrainNearPSNR}{26.07}
\newcommand{\GenOrbitNearPSNR}{25.10}     
\newcommand{\GenPoseSpreadDB}{0.2}        
\newcommand{\GenNearDropDB}{1.3}
\newcommand{\GenOrbitFarSSIM}{0.915}  
\newcommand{\GenOrbitTrainFarSSIM}{0.863}  
\newcommand{\GenOrbitInterpHiSSIM}{0.823}  
\newcommand{\GenOrbitTrainMidSSIM}{0.769}  
\newcommand{\GenOrbitInterpLoSSIM}{0.727}  
\newcommand{\GenOrbitTrainNearSSIM}{0.685}  
\newcommand{\GenOrbitNearSSIM}{0.588}  
\newcommand{\GenRadTwoPSNR}{21.85}\newcommand{\GenRadTwoHalfPSNR}{20.67}
\newcommand{\GenOpaTwoPSNR}{22.30}\newcommand{\GenOpaTwoFourPSNR}{21.33}
\newcommand{\PrepVtpMin}{6.85}            
\newcommand{\SzCompressSec}{11.7}\newcommand{\SzDecompressSec}{7.6}
\newcommand{\LcpCompressSec}{22.3}\newcommand{\LcpDecompressSec}{4.0}
\newcommand{\ParaViewVer}{6.0.1}          

\def\BibTeX{{\rm B\kern-.05em{\sc i\kern-.025em b}\kern-.08em
    T\kern-.1667em\lower.7ex\hbox{E}\kern-.125emX}}

\begin{document}

\title{3D Gaussian Splatting for Scientific Particle Data Compression and Rendering}

\author{%
\IEEEauthorblockN{Bo Jiang\IEEEauthorrefmark{1},
Youyuan Liu\IEEEauthorrefmark{1},
Taolue Yang\IEEEauthorrefmark{1},
Sheng Di\IEEEauthorrefmark{2},
Sian Jin\IEEEauthorrefmark{1}\thanks{Corresponding author: Sian Jin (sian.jin@temple.edu).}}
\IEEEauthorblockA{\IEEEauthorrefmark{1}Temple University, Philadelphia, PA, USA \\
\{bo.jiang, youyuan.liu, taolue.yang, sian.jin\}@temple.edu}
\IEEEauthorblockA{\IEEEauthorrefmark{2}Argonne National Laboratory, Lemont, IL, USA \\
sdi1@anl.gov}
}

\maketitle

\thispagestyle{fancy}
\lhead{}\chead{}\rhead{}
\cfoot{}\rfoot{}
\lfoot{\footnotesize SC26, November 15-20, 2026, Chicago, Illinois, USA\newline 979-8-3195-4789-7/26/\$31.00 \copyright 2026 IEEE}
\renewcommand{\headrulewidth}{0pt}
\renewcommand{\footrulewidth}{0pt}

\begin{abstract}
Large-scale particle simulations produce hundreds of millions of particles,
straining storage, transfer, and interactive visualization.
Existing lossy compressors such as SZ3 operate in data space and provide
no guarantees on downstream visualization fidelity.
We propose \emph{ParticleGS}, a visualization-aware framework based on
3D Gaussian Splatting (3DGS) that learns a compact representation
directly optimized for rendered image quality, combining
(1)~a multi-stage, multi-orbit training pipeline,
(2)~VizMapper, a lightweight network that adapts a single trained model
to user-specified visualization parameters at inference time, and
(3)~spatial block training with KD-tree decomposition and global fine-tuning.
On a \HaccParticlesM{}-million-particle HACC cosmological simulation,
our 8-block model reaches \EbPSNR{}\,dB PSNR at ${\EbCR\times}$
compression, outperforming SZ3 by \GainSzLow{}--\GainSzHigh{}\,dB at
comparable ratios, and generalizes without tuning to
additional HACC regions and a dark-matter-only FIRE-2 simulation.
It renders at \GsFpsOneK{}\,FPS on a single GPU,
over ${\SpeedupRounded\times}$ faster than ParaView on the full particle data.

\end{abstract}


\section{Introduction}
\label{sec:intro}

Particle-based simulations are a workhorse of modern computational science.
Cosmological N-body codes such as HACC~\cite{habib2016hacc} track the
gravitational evolution of dark matter through billions of tracer particles;
galaxy-formation projects such as FIRE-2~\cite{hopkins2018fire,wetzel2023public}
release hydrodynamic zoom-in runs alongside dark-matter-only cosmological
volumes with hundreds of millions of particles;
and molecular dynamics codes such as LAMMPS~\cite{thompson2022lammps}
routinely generate trajectories with comparable particle counts.
At the largest scales, individual HACC simulations track trillions of
particles and produce snapshots in the tens to hundreds of terabytes
each, with multi-snapshot campaigns reaching the petabyte range.
Storing, transferring, and interactively exploring datasets at these
scales remains a bottleneck in scientific workflows.

\begin{figure}[t]
  \centering
  \includegraphics[width=\columnwidth]{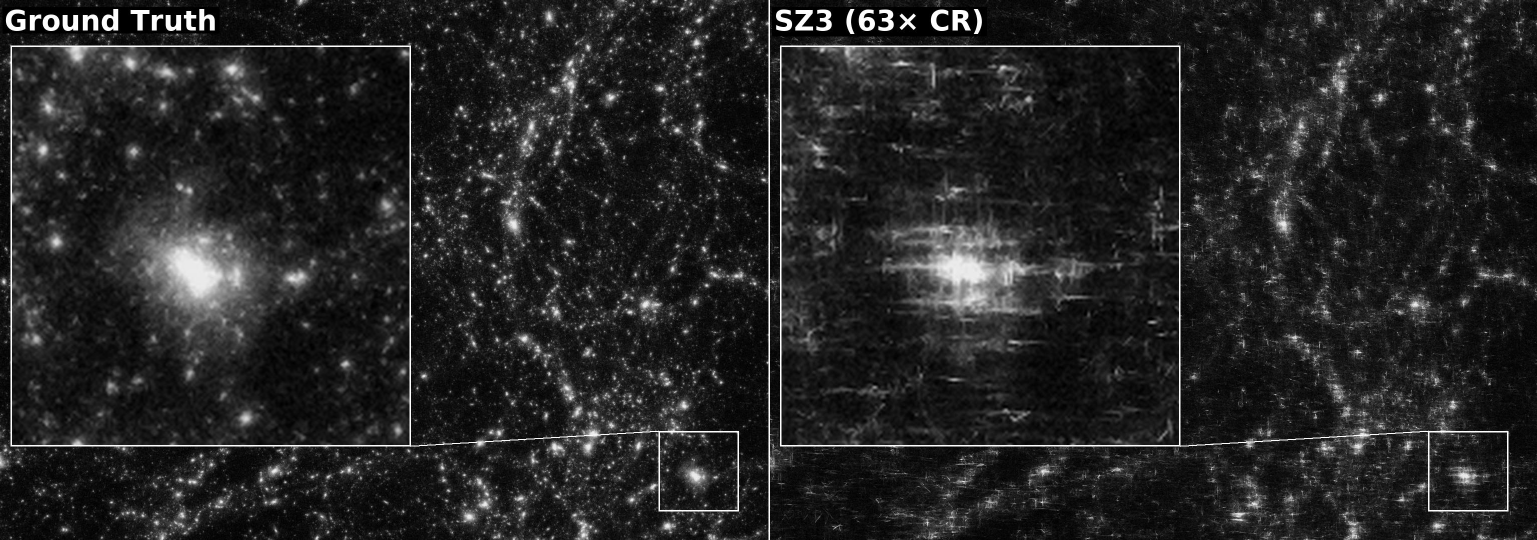}
  \caption{Close-up rendering of \HaccParticlesM{}M HACC particles.
  SZ3 at $\ArtifactSzCR\times$ compression introduces axis-aligned streaking
  artifacts from independent per-axis coordinate quantization,
  destroying the filamentary structure visible in the ground truth.}
  \label{fig:artifact}
  \vspace{-1.5em}
\end{figure}

Interactive visualization of such particle datasets is already
challenging even without compression.
Software such as ParaView~\cite{ahrens2005paraview} renders point clouds by converting each
particle into a screen-space Gaussian splat, a process that scales
linearly with particle count.
For a \HaccParticlesM{}-million-particle dataset, ParaView achieves only \PvFps{}\,FPS
(\PvSecPerFrame{}\,s per frame) when generating our evaluation dataset at $\EvalRes$ resolution
(\cref{tab:perf}), far below the interactive threshold.
Moreover, scientific analysis requires exploring visualization
parameters: users adjust particle radius and opacity to reveal
structures at different scales, from the cosmic web at large radii
to individual filaments at small radii.
Each parameter change triggers a full pipeline re-execution over the
entire particle dataset, making interactive exploration impractical
at scale.

Lossy compression is the standard remedy for reducing data volume,
but existing scientific compressors operate entirely in
\emph{data space}: they minimize reconstruction error of individual
coordinate values with little awareness of how the data will
ultimately be \emph{visualized}.
General-purpose compressors such as
SZ~\cite{di2016fast, liang2022sz3}, ZFP~\cite{lindstrom2014fixed}, and
MGARD~\cite{gong2023mgard} achieve high compression ratios by
exploiting numerical regularity in floating-point arrays and provide
configurable point-wise error bounds, yet these bounds control
data-space distortion, not visual fidelity.
Moreover, these compressors were originally designed for structured
floating-point arrays from grid-based simulations, not for unstructured
particle clouds, and their rate-distortion on particle data is
correspondingly poor.

This disconnect creates two problems in practice.
First, data-space error metrics do not predict visualization quality.
A compressor may report low root-mean-square error while producing
rendered images with clearly visible artifacts, because small
coordinate perturbations can shift particles across density boundaries
or break the local spatial coherence that the renderer relies on.
In our evaluation (\cref{sec:exp:rd}), SZ3 compressed to
${\sim}\SzSizeAtEbCR$\,MB ($\EbCR\times$ ratio) achieves only ${\sim}\SzPsnrAtEbCR$\,dB
PSNR; at ${\sim}\SzSizeAtSbCR$\,MB ($\SbCR\times$) it drops to
${\sim}\SzPsnrAtSbCR$\,dB: poor visual fidelity despite the data-space error
bound being satisfied.
Second, axis-independent compression introduces \emph{directional
artifacts}: SZ and ZFP compress each coordinate axis ($x$, $y$, $z$)
as a separate one-dimensional array, so quantization errors along
different axes are uncorrelated.
When the compressed particles are rendered, this manifests as
directional streaking aligned with the coordinate axes
(\cref{fig:artifact}), an artifact class that is absent from the
original data and that tightening the error bound reduces only
indirectly, at the cost of much lower compression ratios.
Particle-specific compressors such as LCP~\cite{zhang2025lcp} mitigate
axis-independence by reordering particles to exploit spatial locality,
achieving higher compression ratios at the same error bound
(e.g., $\LcpCrAtEbEight\times$ vs.\ SZ3's $\SzCrAtEbEight\times$ at $\varepsilon{=}0.08$).
However, because lossy quantization is then applied within the reordered
stream, its errors become spatially coherent and introduce structured
redistribution artifacts: at $\EbCR\times$ compression, LCP reaches only ${\sim}\LcpPsnrAtEbCR$\,dB, \emph{worse}
than SZ3's ${\sim}\SzPsnrAtEbCR$\,dB at the same size (\cref{tab:rd}).

Neural implicit representations have been explored for scientific
data compression, but they target structured grid data and cannot
handle unstructured particles with adjustable visualization parameters.
Coordinate-based networks such as NeurComp~\cite{lu2021compressive} and
CoordNet~\cite{han2022coordnet} compress volumetric fields, and
NeRF-based methods~\cite{mildenhall2021nerf} have been adapted for
volume rendering of scientific
data~\cite{weiss2022fast, yao2025visnerf},
yet all assume fixed rendering parameters on structured
grids~\cite{kindlmann2014algebraic}.

3D Gaussian Splatting (3DGS)~\cite{kerbl3Dgaussians} offers a compelling
alternative that directly optimizes for rendered image quality.
By representing a scene as a set of anisotropic 3D Gaussians optimized
end-to-end via differentiable rasterization, 3DGS achieves real-time
rendering with high visual fidelity.
The representation is inherently compact: a few hundred thousand
Gaussians can encode the visual appearance of hundreds of millions
of particles, making it naturally aligned with the goal of preserving
visualization fidelity over data-space accuracy.

In this work, we ask: \emph{can 3DGS serve as a visualization-aware
compressed representation for large-scale scientific particle data?}
Adapting 3DGS to this domain, however, introduces challenges not
present in natural-scene reconstruction:

\begin{itemize}
  \item \textbf{Adjustable visualization parameters.}
  Scientific users routinely change particle radius and opacity to reveal
  different structural features.
  Standard 3DGS assumes a fixed scene appearance; a na\"ive approach
  would require retraining for every parameter setting.

  \item \textbf{Multi-scale viewing.}
  Analysts explore data from far-away overview perspectives to close-up
  internal views.
  Training at a single distance leaves the model unable to resolve
  features at other scales.

  \item \textbf{Scale of data.}
  With hundreds of millions of particles, a single model
  lacks the capacity to capture fine-grained spatial structures throughout
  the entire volume.
\end{itemize}

We address these challenges with \emph{ParticleGS}, a
visualization-aware 3DGS pipeline for scientific particle data,
contributing:

\begin{enumerate}
  \item \textbf{A multi-stage/resolution/orbit training pipeline}
  designed for scientific particle rendering.
  We train across multiple camera distances (far, mid, near orbits)
  and progressively increase resolution, enabling the model to learn
  both global structure and local detail (\cref{sec:method:training}).

  \item \textbf{VizMapper}, a lightweight neural network (${\sim}\VmParamsApprox$ parameters)
  that takes per-Gaussian attributes and global visualization parameters
  as input and produces multiplicative corrections to Gaussian.
  This allows a single trained model to adapt to different visualization settings at inference time without retraining (\cref{sec:method:vizmapper}).

  \item \textbf{Spatial block training with global fine-tuning.}
  We partition the particle volume via KD-tree decomposition, train
  independent per-block models in parallel, merge them, and apply a
  global fine-tuning stage.
  This approach increases model capacity where it is most needed and
  improves quality by up to \BlockGainMax{}\,dB over the single-block baseline
  (\cref{sec:method:block}).
\end{enumerate}

We emphasize that ParticleGS is deliberately a \emph{visualization-oriented}
compressed representation: it optimizes the fidelity of rendered images
under interactive visualization parameters and camera motion, rather than
the faithful reconstruction of individual particle positions, attributes,
or identifiers that a general-purpose data compressor targets.
We make this scope explicit because it determines which downstream
analyses the representation supports (\cref{sec:exp:recovery}).

Our pipeline achieves high visual fidelity at large compression ratios
and generalizes across datasets without hyperparameter tuning.
On a \HaccParticlesM{}-million-particle HACC cosmological dataset (\HaccRawGB{}\,GB),
our single-block model achieves \SbPSNR{}\,dB PSNR at ${\SbCR\times}$
compression, while our 8-block model reaches \EbPSNR{}\,dB at
${\EbCR\times}$ compression, outperforming SZ3 by \GainSzLow{}--\GainSzHigh{}\,dB at matched
ratios.
Tested on additional HACC regions and a \FireTwoParticlesM{}M-particle
dark-matter-only FIRE-2 cosmological box (\FireTwoRawGB{}\,GB), the pipeline achieves
consistent quality (\GenPsnrLow{}--\GenPsnrHigh{}\,dB at ${\GenCRLow}$--${\GenCRHigh\times}$).
The 3DGS representation renders at \GsFpsOneK{}\,FPS including VizMapper overhead,
over ${\SpeedupRounded\times}$ faster than ParaView on the full
\HaccParticlesM{}M-particle dataset (\PvFps{}\,FPS), with a ${\EbCR\times}$ smaller footprint.
As a byproduct, the learned Gaussian mixture supports \emph{approximate}
particle recovery via GMM sampling, preserving large-scale density
statistics (correlation $>\RecCorrThreshold$) but not individual
particle positions.

\section{Background}
\label{sec:related}

\subsection{Particle Scientific Data}
\label{sec:bg:particle}

Particle-based simulations model physical systems as collections of
discrete tracer elements, each carrying a three-dimensional spatial
coordinate and, optionally, physical attributes such as velocity or
mass.
Cosmological N-body codes such as HACC~\cite{habib2016hacc} and
Gadget-4~\cite{springel2021simulating} evolve
dark-matter particles under gravitational interactions,
hydrodynamic galaxy-formation simulations like
FIRE-2~\cite{hopkins2018fire} evolve gas and star particles
alongside dark matter, and molecular dynamics simulations such as
LAMMPS~\cite{thompson2022lammps} track atomic
trajectories, all at scales ranging from millions to billions of
particles per snapshot.
Unlike grid-based simulation outputs, where field values reside on a
regular lattice, particle positions form an \emph{unstructured} point
cloud in $\mathbb{R}^3$.
This distinction is consequential: compression and rendering techniques
designed for structured grids (including most neural implicit
representations) do not directly apply to particle data.

Visualization tools such as ParaView~\cite{ahrens2005paraview} render particle datasets by
drawing each particle as a screen-space Gaussian splat, a
semi-transparent sprite with Gaussian falloff parameterized by
a user-specified radius~$r$ and opacity~$\alpha$.
Adjusting these reveals structures at different scales: large
radii with low opacity expose large-scale filamentary networks,
while small radii with higher opacity resolve individual density
features. This parameter sensitivity makes interactive control
essential, so any compressed representation must accommodate
varying visualization settings rather than a fixed appearance.

\subsection{Lossy Compression for Particle Data}
\label{sec:bg:compression}

Lossy scientific compressors reduce floating-point data volumes by
tolerating bounded reconstruction error, enabling storage and transfer
of datasets that would otherwise be prohibitively large.
SZ~\cite{di2016fast} and its successor SZ3~\cite{liang2022sz3} predict
each value from its neighbors and quantize the prediction residual;
a user-specified absolute error bound~$\varepsilon$ guarantees that
no reconstructed value deviates from the original by more than
$\varepsilon$.
ZFP~\cite{lindstrom2014fixed} partitions data into small blocks, applies
a near-orthogonal transform, and encodes the coefficients at a
configurable bit rate.
MGARD~\cite{gong2023mgard} provides rigorous $L^\infty$ error
guarantees through multiresolution decomposition.
TTHRESH~\cite{ballester2019tthresh} decomposes data tensors via
higher-order SVD with optimal quantization.
More recently, HPEZ~\cite{liu2024high} combines multi-component
interpolation with auto-tuning to achieve state-of-the-art
rate-distortion among prediction-based compressors.
GPU-accelerated variants such as cuSZ~\cite{tian2020cusz} and
cuSZ-i~\cite{liu2024cuszi} bring error-bounded compression to
massively parallel hardware, achieving throughputs of tens of GB/s
on modern GPUs.
Tao et~al.~\cite{tao2017depth} systematically evaluated these
compressors on N-body simulation data, showing that the low spatial
coherence inherent in particle coordinates limits the compression
ratios achievable with prediction-based methods.
These compressors were designed for structured grid fields; when applied
to particle coordinates, they treat each axis ($x$, $y$, $z$) as a
separate one-dimensional array and compress the three arrays
independently.
As a result, quantization errors along different axes are uncorrelated,
and the error bounds control data-space distortion of individual
values rather than the fidelity of downstream
visualizations, a disconnect whose consequences we quantify
in \cref{sec:exp:rd}.

Recent work has begun to address this disconnect.
Liu et~al.~\cite{liu2022dynamic} allow users to specify target
quality metrics (PSNR, SSIM) rather than pointwise error bounds,
auto-tuning compression parameters to meet the target.
Jiao et~al.~\cite{jiao2022toward} develop theory and algorithms for
preserving downstream quantities of interest (spectra, histograms)
during error-bounded compression.
On the learning side, AE-SZ~\cite{liu2021exploring} combines autoencoders
with SZ-style error bounding, achieving significantly higher
compression ratios at the cost of throughput.
However, all these methods still operate in data space and do not
directly optimize for rendered image quality.
Z-checker~\cite{tao2019z} provides a standard framework for
assessing lossy compression quality, but its metrics remain
data-centric.

Particle-specific compressors mitigate per-axis artifacts but do not
close the gap between data-space error and visualization quality.
LCP~\cite{zhang2025lcp} reorders particles along a space-filling
curve before applying SZ-style prediction and quantization, grouping
spatially nearby particles to exploit local coordinate similarity
and achieve higher compression at the same error bound.
Reordering itself is render-invariant (particle order does not affect
the rendered density field), but the subsequent lossy quantization is
applied within the reordered stream, so its reconstruction errors become
spatially coherent along the space-filling curve, displacing particles
into structured artifacts rather than incoherent per-point noise.

\subsection{3D Gaussian Splatting}
\label{sec:bg:3dgs}

Neural Radiance Fields (NeRF)~\cite{mildenhall2021nerf} sparked rapid
progress in neural scene representations; subsequent hash-based
encodings~\cite{muller2022instant} greatly accelerated both
training and rendering.
3D Gaussian Splatting (3DGS)~\cite{kerbl3Dgaussians} represents a scene
as a set of anisotropic 3D Gaussians, each characterized by a center
position, a covariance matrix (stored compactly as a scale vector and
rotation quaternion), an opacity value, and spherical-harmonic color
coefficients for view-dependent appearance.
To render an image, each Gaussian is projected onto the image plane as
a 2D ellipse, and overlapping contributions are composited
front-to-back via alpha blending, analogous to the splatting
operation used in scientific visualization tools but with learned
rather than prescribed parameters.
A tile-based CUDA rasterizer sorts Gaussians per screen tile and
performs this compositing in parallel, enabling real-time rendering at
hundreds of frames per second.
The rasterizer is fully differentiable, so an image-space loss
backpropagates gradients to every Gaussian parameter, allowing the
entire representation to be optimized end-to-end from multi-view
images without explicit 3D supervision.
An adaptive density control scheme periodically splits Gaussians in
under-reconstructed regions and prunes those with negligible opacity,
automatically adjusting model capacity to match scene complexity.
Compared to NeRF, 3DGS offers several advantages: the explicit
Gaussian representation avoids costly per-ray sampling, achieving
$100$--$300\times$ faster rendering; training converges in minutes
rather than hours; and the model can be stored and streamed as a
compact set of per-Gaussian parameters.

Existing 3DGS methods, however, all target natural-scene
reconstruction from photographs, where scene appearance is fixed at
capture time.
Subsequent work has improved alias-free rendering via
prefiltering~\cite{yu2024mip}, reduced storage through pruning and
quantization~\cite{lee2024compact, fang2024mini},
extended the framework to dynamic
scenes~\cite{wu20244d}, and scaled to large environments
through spatial partitioning~\cite{lin2024vastgaussian};
none of these address scientific visualization.
Scientific particle visualization differs in two key respects:
users routinely change visualization parameters ($r$, $\alpha$) to
reveal different structural features, and the input is an unstructured
point cloud rather than a set of calibrated photographs.
A standard 3DGS model trained under one parameter setting cannot
generalize to others without retraining, motivating the extensions
described in \cref{sec:method}.

\section{Methodology}
\label{sec:method}

\begin{figure*}[!t]
  \centering
  \includegraphics[width=\textwidth]{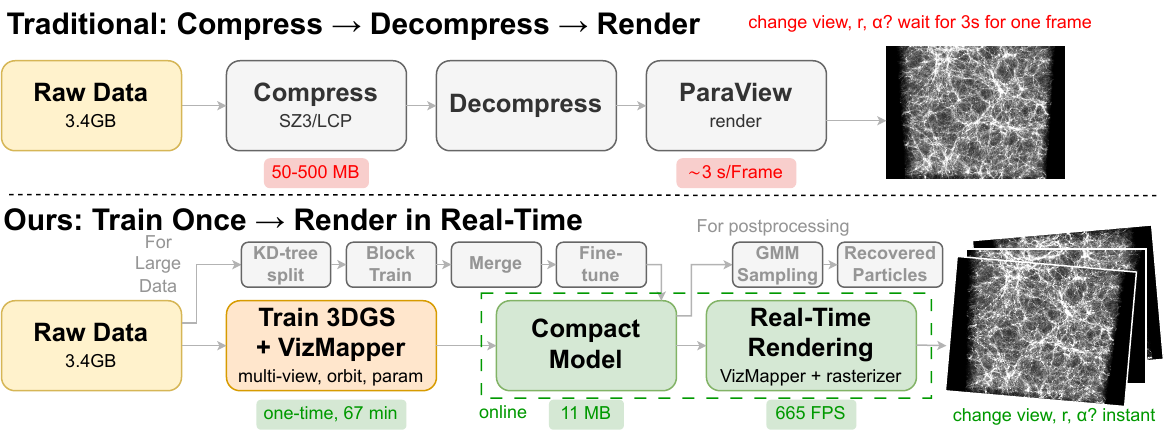}
  \caption{Comparison of the traditional compress--decompress--render
  pipeline~(top) and our ParticleGS pipeline~(bottom).
  The traditional workflow requires lossy compression, full decompression,
  and slow ParaView rendering (${\sim}\PvSecPerFrame$\,s/frame) for every view change.
  Our approach trains a compact 3DGS model with VizMapper once
  (\TrainTimeMin{}\,min offline), then delivers real-time interactive rendering
  at \GsFpsOneK{}\,FPS with on-the-fly parameter control.
  For large datasets, spatial block training partitions the volume
  via KD-tree (\cref{sec:method:block}); as a byproduct, the learned
  Gaussians enable particle recovery via GMM sampling (\cref{sec:method:recovery}).}
  \label{fig:pipeline}
\end{figure*}

Our goal is to learn a compact 3DGS representation that faithfully
reproduces ParaView renderings of a large particle dataset across
diverse viewpoints and visualization parameters, at high compression
and real-time rendering (\cref{fig:pipeline}).
We formulate the task as optimizing 3D Gaussians whose rasterized
images match ParaView renderings (\cref{sec:method:problem}).
VizMapper, a lightweight MLP, modulates Gaussian scale and opacity
to adapt to user-specified visualization parameters at inference
time (\cref{sec:method:vizmapper}).
A multi-orbit, multi-resolution schedule with content-masked L1 loss
trains the model end-to-end (\cref{sec:method:training}).
For large datasets, we partition the volume via KD-tree, train
per-block models in parallel, merge them, and globally fine-tune
to repair seams (\cref{sec:method:block}).
As a byproduct, the learned Gaussian mixture serves as a density
model from which particle positions can be recovered via GMM
sampling (\cref{sec:method:recovery}).

\subsection{Problem Formulation}
\label{sec:method:problem}

Let $\mathcal{P} = \{\mathbf{p}_i\}_{i=1}^{N}$ denote $N$ particle positions
$\mathbf{p}_i \in \mathbb{R}^3$. ParaView renders $\mathcal{P}$ into an image
$I = \mathcal{R}(\mathcal{P}; r, \alpha, \mathbf{c})$ via its Point Gaussian
representation, where $r$, $\alpha$, and $\mathbf{c}$ are the particle
radius, opacity, and camera pose.
Our goal is to learn a 3DGS representation
$\mathcal{G} = \{(\boldsymbol{\mu}_j, \boldsymbol{\Sigma}_j, o_j, \mathbf{f}_j)\}_{j=1}^{M}$
with $M \ll N$ Gaussians, such that
$\hat{I} = \mathcal{R}_{\text{3DGS}}(\mathcal{G}; r, \alpha, \mathbf{c})$
approximates $I$ across viewpoints and visualization parameters.
Here $\boldsymbol{\mu}_j$ is the Gaussian center, $\boldsymbol{\Sigma}_j$ the
covariance (parameterized via scaling $\mathbf{s}_j$ and rotation $\mathbf{q}_j$),
$o_j$ the opacity, and $\mathbf{f}_j$ the color features.

\subsection{VizMapper: Visualization Parameter Adaptation}
\label{sec:method:vizmapper}

We introduce VizMapper, a lightweight MLP that modulates each
Gaussian's scale and opacity conditioned on user-specified
visualization parameters, enabling one model to generalize
across rendering settings without retraining.

\begin{figure}[t]
  \centering
  \includegraphics[width=\columnwidth]{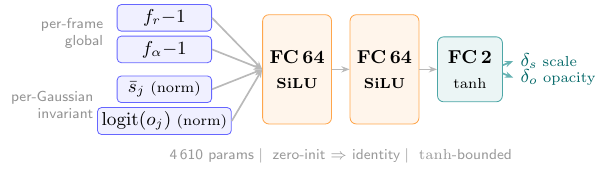}
  \caption{VizMapper architecture. A lightweight MLP maps per-Gaussian
  attributes and global visualization parameters to scale and opacity
  corrections.}
  \label{fig:vizmapper}
\end{figure}

Na\"ive approaches to handling varying visualization parameters are
either too expensive or sacrifice rendering quality; VizMapper
addresses both limitations.
In scientific visualization, users frequently adjust the particle radius $r$
and opacity $\alpha$ to reveal different structural features.
One na\"ive approach is to train a separate model for each parameter
combination, which is prohibitively expensive and cannot generalize to
unseen settings.
Another is to train on randomized visualization parameters without
adaptation, which forces the Gaussians to represent an average over all
settings, resulting in blurry renderings.
VizMapper solves this by learning a lightweight per-Gaussian correction
that adapts scale and opacity to the current visualization setting
(\cref{fig:vizmapper}).

VizMapper is a 3-layer MLP $\phi_\theta$ ($\VmArch$,
\VmParams{} parameters) with SiLU activations.
It modulates only scale and opacity (not position or color) because
changing $r$ and $\alpha$ affects only particle size and transparency.
During training, each image is rendered with a randomly sampled
radius~$r$ and opacity~$\alpha$
(\cref{sec:method:training}).
Each dataset has default values $r_0$ and $\alpha_0$;
we express the current settings as ratios
$f_r = r / r_0$ and $f_\alpha = \alpha / \alpha_0$,
so that $f = 1$ means ``use the default'' and, e.g., $f_r = 1.5$
means ``50\% larger than the default radius.''
VizMapper takes as input the relative change from the default
($f - 1$, which is zero when no adaptation is needed) together with
two per-Gaussian attributes, the mean log-scale $\bar{s}_j$ and
the opacity logit $\operatorname{logit}(o_j)$, that let the network
produce different corrections for different Gaussians:
\begin{equation}
  \mathbf{x}_j = \left[
    f_r - 1, \;\;
    f_\alpha - 1, \;\;
    \frac{\bar{s}_j - \mu_s}{\Delta_s}, \;\;
    \frac{\operatorname{logit}(o_j) - \mu_o}{\Delta_o}
  \right].
\end{equation}
The per-Gaussian log-scale and opacity logit (the same quantities
stored internally by 3DGS) are affine-normalized by constants
$(\mu_s, \Delta_s)$ and $(\mu_o, \Delta_o)$ so that all four inputs
span approximately $[-1, 1]$.
These constants are estimated once from a pilot run on a reference
dataset and fixed for all subsequent experiments
(\cref{sec:exp}).
Including per-Gaussian attributes is essential: each Gaussian
represents a different local particle structure, and the
correction depends on its own scale and opacity.

The network outputs two $\tanh$-bounded residual corrections:
\begin{equation}
  (\delta_s, \delta_o) = \phi_\theta(\mathbf{x}_j), \;\;
  \delta_s \in [-\delta_s^{\max}, \delta_s^{\max}], \;\;
  \delta_o \in [-\delta_o^{\max}, \delta_o^{\max}].
\end{equation}
Both the adapted scale and opacity follow the same three-factor
structure: base value $\times$ uniform factor $\times$ per-Gaussian
correction.
For scale, this is a direct product of the Gaussian's intrinsic
scale~$\mathbf{s}_j$, the global rescaling~$f_r$, and the learned
residual~$(1 + \delta_s)$:
\begin{equation}
  \mathbf{s}_j' = \mathbf{s}_j \cdot f_r \cdot (1 + \delta_s).
\end{equation}
Opacity has the same multiplicative decomposition (base $o_j$,
uniform factor $f_\alpha$, residual $(1 + \delta_o)$), but since
3DGS already stores opacity as a logit and applies $\sigma$ during
rendering, we inject the correction directly into the logit before
the existing sigmoid:
\begin{equation}
  o_j' = \sigma\!\bigl(\operatorname{logit}(o_j)
    + \log \max(f_\alpha (1{+}\delta_o),\;\epsilon)\bigr),
\end{equation}
where $\sigma$ is the sigmoid function and $\epsilon$ is a lower bound
on the multiplicative opacity factor (preventing the network from driving
opacities arbitrarily close to zero).

VizMapper is trained jointly with all 3DGS parameters end-to-end.
The output layer is zero-initialized so that $\delta_s = \delta_o = 0$
at the start of training, ensuring the network begins as the identity
mapping and does not interfere with early geometric optimization.
The network is kept small (\VmParams{} parameters) since it runs
once per Gaussian per frame. For initial training we use
$\delta_s^{\max} = \DeltaSMax$, $\delta_o^{\max} = \DeltaOMax$, and $\epsilon = \InitEpsilon$;
these ranges widen during fine-tuning (\cref{sec:method:block}).

\subsection{Training Pipeline}
\label{sec:method:training}

Our training pipeline normalizes particle coordinates, samples
visualization parameters from Beta distributions, and follows a
two-stage, multi-resolution schedule that progressively increases
image resolution across three camera orbits.

We first normalize particle coordinates and construct an initial
point cloud.
Scientific particle coordinates span large ranges
(e.g., $[0, \HaccDomain]^3$ for HACC).
We center and scale the bounding box to fit within $[-1,1]^3$,
recording the transform in a shared \texttt{normalization.json}.
From the normalized volume we subsample ${\sim}\InitPoints$ particles
as a colored point cloud (PLY format) to initialize the 3DGS
optimization.
This subsample seeds optimization \emph{only}: the ground-truth images
that supervise training are rendered from the \emph{full} set of all
\HaccParticlesM{}M particles, and adaptive densification subsequently
expands the model to \SbGaussians{} (single-block) or \EbGaussians{}
(8-block) Gaussians.
The model is never trained to reproduce the visualization from the
${\sim}\InitPoints$-particle subset alone.

We pre-sample visualization parameters per training image so that
the model learns to handle a range of rendering settings.
Specifically, the particle radius~$r$ and opacity~$\alpha$
are independently drawn from Beta distributions centered on their
defaults ($r_0 = \DefaultR$, $\alpha_0 = \DefaultAlpha$):
$r \sim r_{\min} + (r_{\max} - r_{\min})\,\text{Beta}(\beta, \beta)$,
with concentration $\beta = \BetaConc$,
$r \in [\RadiusMin, \RadiusMax]$, and $\alpha \in [\OpacityMin, \OpacityMax]$.
The Beta shape concentrates samples near the defaults while still
covering extreme settings, so the model devotes most capacity to
the most commonly used parameter combinations.
VizMapper (\cref{sec:method:vizmapper}) receives the corresponding
factors $f_r$ and $f_\alpha$ each iteration.

\begin{figure}[!t]
  \centering
  \includegraphics[width=\columnwidth]{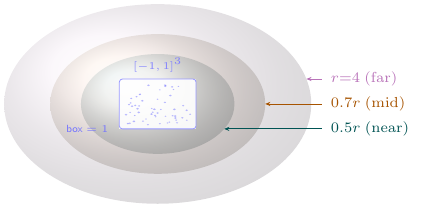}
  \caption{Multi-distance camera orbits.
  Three families of orbital camera trajectories at far ($1.0\times$),
  mid ($0.7\times$), and near ($0.5\times$) base radius surround the
  normalized particle volume $[-1,1]^3$, providing both distant
  overviews and close-up views.}
  \label{fig:camera_orbits}
  \vspace{-1.5em}
\end{figure}

We use multi-distance camera orbits and a two-stage progressive
resolution schedule.
Scientific exploration requires both distant overviews and close-up
views, so we generate three families of orbital camera trajectories at
far ($1.0\times$), mid ($0.7\times$), and near ($0.5\times$)
base radius from the scene center (\cref{fig:camera_orbits}),
each with uniform azimuth and elevation angles.

Training proceeds in two resolution stages that share the multi-orbit
camera distribution.
Stage~1 trains at $\StageOneRes$ for \StageOneIters{} iterations
with adaptive densification, establishing the coarse Gaussian
distribution at low computational cost.
Stage~2 continues at full $\StageTwoRes$ resolution for
\StageTwoIters{} iterations, refining high-frequency structure with
gradually tapering densification.
The first \StageTwoExtIters{} iterations of stage~2 use the same external
orbits as stage~1; the remaining \StageTwoMixIters{} iterations additionally
mix in a small fraction (\InternalMixPct\%) of cameras placed inside the
particle volume, which improves robustness for close-up internal views
without materially affecting the rendered metric on external orbits.
Total training: \TotalIters{} iterations (${\sim}\TrainTimeMin$ minutes on a
single GPU).

We use a pure L1 loss between predicted and ground-truth images,
restricted to a content mask $\mathcal{M}$ that excludes the
black background:
\begin{equation}
  \mathcal{L} = \frac{1}{|\mathcal{M}|} \sum_{(u,v) \in \mathcal{M}}
  \left| \hat{I}(u,v) - I(u,v) \right|.
  \label{eq:loss}
\end{equation}
Adding DSSIM, standard in natural-scene 3DGS, inflates model size
by $\DssimSizeIncreasePct\%$ without improving masked PSNR (\cref{sec:exp:ablation}),
since the SSIM signal is dominated by background.
SH degree~0 throughout saves another \ShSavingsPct\% versus degree~2,
since particle rendering is view-independent.

\subsection{Spatial Block Training with Global Fine-Tuning}
\label{sec:method:block}

\begin{figure}[t]
  \centering
  \includegraphics[width=\columnwidth]{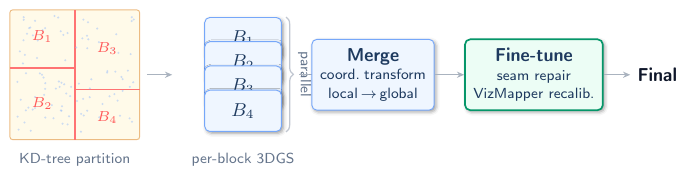}
  \caption{Spatial block training pipeline.
  A KD-tree partitions the particle volume into $K$ blocks, each
  trained independently in parallel.
  Block Gaussians are merged into a single model via coordinate
  transform (local\,$\to$\,global), then globally fine-tuned to
  repair boundary seams and recalibrate VizMapper.}
  \label{fig:block_training}
  \vspace{-1.5em}
\end{figure}

For datasets that exceed a single model's capacity, we partition the
volume into blocks via KD-tree, train each block independently, merge
them into a unified representation, and fine-tune globally to repair
boundary seams.

We partition the volume into balanced blocks so that each can be
trained independently.
A single 3DGS model has limited capacity to represent fine-grained
structure across a large particle volume, and training cost grows
with scene complexity.
We therefore partition the volume into $K$ blocks using a KD-tree that
recursively bisects along the axis of greatest extent (\cref{fig:block_training}),
producing spatially compact blocks with roughly equal particle
counts so that per-block training cost is balanced.
Each block is trained independently using the pipeline of
\cref{sec:method:training}, with camera orbits centered on each
block's local bounding box via per-block normalization.
Blocks can be trained in parallel across multiple GPUs.

After per-block training, we transform each block's Gaussians from
their local coordinate frame back to the global frame.
Positions are mapped via
$\mathbf{p}_{\text{global}} = \mathbf{p}_{\text{local}}
\cdot (s_{\text{global}} / s_{\text{block}})
+ (\mathbf{c}_{\text{block}} - \mathbf{c}_{\text{global}})
  \cdot s_{\text{global}}$,
where $s$ denotes the normalization scale factor and $\mathbf{c}$ the
coordinate-frame center;
log-scale parameters are shifted by
$\log(s_{\text{global}} / s_{\text{block}})$.
All block Gaussians are merged into one PLY.

The merged model is then fine-tuned end-to-end on the original
multi-orbit training data at $\EvalRes$.
Fine-tuning at reduced resolution suffices because the goal is to
repair block-boundary seams and recalibrate VizMapper for the
combined Gaussian population, not to recover additional
high-frequency detail.
Because each block's VizMapper was trained on only that block's
Gaussian distribution, the merged population is more heterogeneous;
we therefore widen the correction ranges
($\delta_o^{\max} = \FtDeltaOMax$, $\delta_s^{\max} = \FtDeltaSMax$, $\epsilon = \FtEpsilon$)
to give the network enough room to adapt.
Fine-tuning runs for \FtIters{} iterations with mild densification
enabled during the first \FtDensifyUntil{} iterations.

\subsection{Particle Recovery via GMM Sampling}
\label{sec:method:recovery}

As a byproduct, we reinterpret the optimized Gaussians as a
Gaussian Mixture Model and sample from it to recover particle
positions that preserve large-scale density statistics.

Beyond visualization, the trained 3DGS model can serve as a
density model of particle positions.
We reinterpret the optimized Gaussians
$\mathcal{G} = \{(\boldsymbol{\mu}_j, \boldsymbol{\Sigma}_j, o_j)\}_{j=1}^{M}$
as a Gaussian Mixture Model (GMM), where each component's mixing
weight is proportional to its opacity:
\begin{equation}
  w_j = \frac{o_j}{\sum_{k=1}^{M} o_k}.
  \label{eq:gmm_weight}
\end{equation}
To recover $N$ particle positions, we first draw component indices
$\{z_i\}_{i=1}^{N}$ from the categorical distribution
$z_i \sim \text{Cat}(w_1, \dots, w_M)$,
then sample particles from the Gaussian:
\begin{equation}
  \hat{\mathbf{p}}_i = \boldsymbol{\mu}_{z_i}
    + \mathbf{R}_{z_i}\,\text{diag}(\boldsymbol{\sigma}_{z_i})\,\boldsymbol{\epsilon},
  \quad \boldsymbol{\epsilon} \sim \mathcal{N}(\mathbf{0}, \mathbf{I}),
  \label{eq:gmm_sample}
\end{equation}
where $\mathbf{R}_j$ and $\boldsymbol{\sigma}_j$ parameterize
$\boldsymbol{\Sigma}_j$. Sampled positions are then mapped from
the normalized $[-1,1]^3$ frame back to the original coordinates.

Recovering \HaccParticlesM{}M particles takes ${\sim}\RecTimeSec$\,s on one GPU
with no additional training. Because the Gaussians are optimized
for \emph{rendering} fidelity, recovered particles serve as a
statistical proxy, preserving large-scale statistics (power
spectrum, density PDF) but not local structure (\cref{sec:exp:recovery}).

\section{Experiments}
\label{sec:exp}

We evaluate on large-scale particle datasets from multiple simulation codes.
Ablation studies (\cref{sec:exp:ablation}) isolate each design choice;
block training analysis (\cref{sec:exp:block}) and generalization
(\cref{sec:exp:generalization}) validate scalability.
We compare against SZ3 and LCP in rate-distortion (\cref{sec:exp:rd}),
profile rendering and pipeline performance (\cref{sec:exp:perf}),
and assess whether the learned representation preserves spatial
statistics of the original distribution (\cref{sec:exp:recovery}).

\subsection{Dataset and Setup}
\label{sec:exp:setup}

We evaluate on snapshots from two cosmological simulation codes.
Our primary dataset is a \HaccParticlesM{}M-particle HACC N-body
snapshot~\cite{habib2016hacc} over a $[0,\HaccDomain]^3$ domain,
stored as three float32 arrays ($xx$, $yy$, $zz$), totaling \HaccRawGB{}\,GB.
For generalization (\cref{sec:exp:generalization}), we also use
(i)~four ${\sim}\HaccGenParticlesM$M-particle HACC snapshots from
disjoint spatial regions of a separate HACC dataset, and
(ii)~the \FireTwoParticlesM{}M-particle dark-matter-only cosmological box
from the FIRE-2 public data release~\cite{wetzel2023public}
($[0,\!\FireTwoDomain]^3$, \FireTwoRawGB{}\,GB).

Ground-truth images are rendered with ParaView~\ParaViewVer{}'s Point Gaussian
representation at $\EvalRes$ (the same version is used for all timing
comparisons in \cref{sec:exp:perf}). We evaluate at three orbital camera
distances (far: $\OrbitFar\times$, mid: $\OrbitMid\times$, near: $\OrbitNear\times$ base radius),
\EvalFrames{} frames each (\EvalTotalFrames{} total). Per-frame radius and opacity
are sampled from $\text{Beta}(\BetaConc, \BetaConc)$ over $r \in [\RadiusMin, \RadiusMax]$ and
$\alpha \in [\OpacityMin, \OpacityMax]$ (defaults $r_0 = \DefaultR$, $\alpha_0 = \DefaultAlpha$), matching
real scientific exploration.

We report PSNR and model size as primary metrics.
We additionally report SSIM~\cite{wang2004image} and \emph{masked PSNR},
i.e., PSNR computed only over foreground pixels (where either ground truth
or rendering is non-black), excluding the uniform background;
tables report full-image PSNR unless explicitly labeled masked.
We compare against two lossy compressors:
\textbf{SZ3}~\cite{liang2022sz3}, a general-purpose
lossy compressor for scientific data (ABS mode, per-axis compression);
and \textbf{LCP}~\cite{zhang2025lcp}, a particle-specific compressor
that exploits spatial locality through particle reordering and joint
three-axis compression.
Both baselines produce decompressed coordinates rendered through
the identical ParaView pipeline for fair comparison;
we sweep error-bound values to trace full rate-distortion curves.
All experiments run on two NVIDIA RTX PRO 6000 GPUs.

\subsection{Ablation Studies}
\label{sec:exp:ablation}

Unless noted otherwise, the ablations, block-training analysis, and
rate-distortion comparisons (\cref{sec:exp:ablation,sec:exp:block,sec:exp:rd})
all use the primary \HaccParticlesM{}M-particle HACC snapshot of
\cref{sec:exp:setup}.
We ablate two design axes: core training components and fine-tune loss.
VizMapper is the most impactful single factor ($\AblNoVmDelta$\,dB without
it), followed by progressive multi-resolution training ($\AblNoProgDelta$\,dB)
and multi-orbit cameras ($\AblNoOrbitDelta$\,dB).
All three variants in \cref{tab:ablation_core} share the full E25 schedule
(\TotalIters{} iterations, pure L1 + content-mask loss, antialiasing,
SH degree~0, the same densification configuration, and identical
Beta-sampled radius/opacity training images); each row modifies only the
named component.

Disabling VizMapper (\texttt{static\_viz}=true; cameras, schedule, and
densification unchanged) drops PSNR by \AblNoVmDelta{}\,dB
(\cref{tab:ablation_core}). The static model must approximate all
radius/opacity combinations with one fixed set of Gaussian opacities and
scales; densification compensates by inflating the model to
\AblNoVmGaussianPct{}\% more Gaussians
(\AblNoVmGaussians{} vs.\ \SbGaussians{}), yet still scores
\AblNoVmDelta{}\,dB lower: extra capacity cannot substitute for
per-parameter adaptation.
Removing progressive resolution collapses the three stages into a single
\TotalIters{}-iteration stage at $\StageOneRes$ (cameras and densification
unchanged), costing $\AblNoProgDelta$\,dB; coarse-to-fine refinement is
worth more than equal iterations at the lowest resolution alone.
Restricting all stages to the far $1.0\times$ orbit (resolution schedule
and VizMapper unchanged) costs $\AblNoOrbitDelta$\,dB, since the model
never sees mid- or near-distance training views.

\begin{table}[t]
  \centering
  \caption{Core training method ablation.
  Each row disables one component from the default pipeline (\SbPSNR{}\,dB);
  removing any single component degrades quality, with progressive resolution
  contributing the largest share.}
  \label{tab:ablation_core}
  \begin{tabular}{lcccc}
    \toprule
    \textbf{Ablated Component} & \textbf{PSNR} & \textbf{$\Delta$}
    & \textbf{Size} & \textbf{Gaussians} \\
    \midrule
    Full pipeline (default) & \SbPSNR{}\,dB & --- & \SbSizeMB{}\,MB & \SbGaussians{} \\
    \quad w/o VizMapper & \AblNoVmPSNR{}\,dB & $\AblNoVmDelta$ & \AblNoVmSizeMB{}\,MB & \AblNoVmGaussians{} \\
    \quad w/o progressive res. & \AblNoProgPSNR{}\,dB & $\AblNoProgDelta$ & \AblNoProgSizeMB{}\,MB & \AblNoProgGaussians{} \\
    \quad w/o multi-orbit & \AblNoOrbitPSNR{}\,dB & $\AblNoOrbitDelta$ & \AblNoOrbitSizeMB{}\,MB & \AblNoOrbitGaussians{} \\
    \bottomrule
  \end{tabular}
\end{table}

Pure L1 with densification is the best fine-tune recipe
(\cref{tab:ablation_ft}).
Pure DSSIM loss degrades by $\AblFtPureDssimDelta$\,dB with $\AblFtPureDssimBloat\times$ model bloat
(\AblFtPureDssimSizeMB{}\,MB vs.\ \EbSizeMB{}\,MB), because DSSIM's patch-based gradients
encourage many small Gaussians to cover each patch boundary.
Even a 10\% DSSIM blend costs $\AblFtDssimBlendDelta$\,dB.
Disabling densification costs $\AblFtNoDensifyDelta$\,dB, as the merged model needs
new Gaussians to fill gaps between block boundaries.

\begin{table}[t]
  \centering
  \caption{Fine-tune recipe ablation on the 8-block model.
  Production config (pure L1, AA, densification) is the baseline.}
  \label{tab:ablation_ft}
  \setlength{\tabcolsep}{4pt}
  \begin{tabular}{lccc}
    \toprule
    \textbf{Config} & \textbf{PSNR} & \textbf{$\Delta$} & \textbf{Size} \\
    \midrule
    Production recipe & \EbPSNR{}\,dB & --- & \EbSizeMB{}\,MB \\
    \quad w/o densification & \AblFtNoDensifyPSNR{}\,dB & $\AblFtNoDensifyDelta$ & \AblFtNoDensifySizeMB{}\,MB \\
    \quad pure DSSIM & \AblFtPureDssimPSNR{}\,dB & $\AblFtPureDssimDelta$ & \AblFtPureDssimSizeMB{}\,MB \\
    \quad L1 + 0.1 DSSIM & \AblFtDssimBlendPSNR{}\,dB & $\AblFtDssimBlendDelta$ & \AblFtDssimBlendSizeMB{}\,MB \\
    \bottomrule
  \end{tabular}
\end{table}

\subsection{Block Training Analysis}
\label{sec:exp:block}

Spatial block decomposition improves quality by up to $+\BlockGainMax$\,dB
over the single-block baseline, with diminishing returns beyond
8~blocks (\cref{tab:block}).
The largest gain comes from 1$\to$2 blocks (+\BlkTwoDelta{}\,dB after
fine-tuning), as even a coarse partition lets each sub-model
specialize in a smaller spatial region.

\begin{table}[!t]
  \centering
  \caption{Block training sweep.
  ``Merged'' is measured immediately after coordinate-transform concatenation;
  ``Fine-tuned'' applies \FtIters{} iterations of F16 global fine-tuning.
  $\Delta$ is relative to the single-block baseline (\SbPSNR{}\,dB).}
  \label{tab:block}
  \setlength{\tabcolsep}{4pt}
  \begin{tabular}{rcccccc}
    \toprule
    \textbf{Blocks} & \textbf{Merged} & \textbf{Fine-tuned} & \textbf{$\Delta$}
    & \textbf{SSIM} & \textbf{Gaussians} & \textbf{Size} \\
    \midrule
    1 & --- & \SbPSNR & --- & \SbSSIM & \SbGaussians & \SbSizeMB{}\,MB \\
    2  & \BlkTwoMergedPSNR & \BlkTwoFtPSNR & \BlkTwoDelta & \BlkTwoSSIM & \BlkTwoGaussians & \BlkTwoSizeMB{}\,MB \\
    4  & \BlkFourMergedPSNR & \BlkFourFtPSNR & \BlkFourDelta & \BlkFourSSIM & \BlkFourGaussians & \BlkFourSizeMB{}\,MB \\
    8  & \BlkEightMergedPSNR & \BlkEightFtPSNR & \BlkEightDelta & \BlkEightSSIM & \BlkEightGaussians & \BlkEightSizeMB{}\,MB \\
    16 & \BlkSixteenMergedPSNR & \BlkSixteenFtPSNR & \BlkSixteenDelta & \BlkSixteenSSIM & \BlkSixteenGaussians & \BlkSixteenSizeMB{}\,MB \\
    \bottomrule
  \end{tabular}
\end{table}

Fine-tuning is essential: merged quality degrades as the number of
blocks grows beyond two (\BlkTwoMergedPSNR{}\,dB for 2 blocks down to
\BlkSixteenMergedPSNR{}\,dB for 16 blocks), but fine-tuning fully
recovers and surpasses single-block quality in all cases,
confirming that global optimization after merging is necessary.

Rendering quality saturates beyond 4 blocks: doubling from 4 to 8
blocks adds only \BlockSatDelta{}\,dB (\BlkFourFtPSNR{} $\to$ \BlkEightFtPSNR{}\,dB) while increasing
model size by \BlockSizeIncFourToEightPct{}\%.
However, more blocks enable training-time parallelism for larger
datasets and improve the fidelity of recovered particle data
(\cref{tab:recovery_blocks}).

\begin{figure}[!t]
  \centering
  \includegraphics[width=\columnwidth]{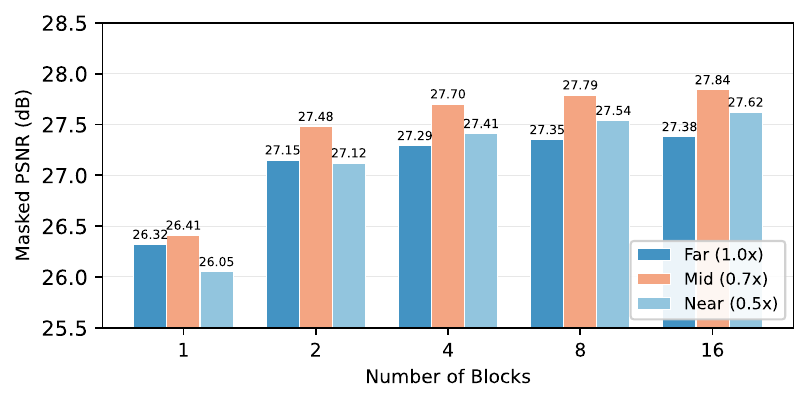}
  \caption{Fine-tuned masked PSNR by viewing distance for different block
  counts. More blocks consistently improve quality across all distances,
  with the largest gain from 1 to 2 blocks.}
  \label{fig:block_distance}
\end{figure}

Block training primarily benefits mid and near views
(\cref{fig:block_distance}).
For the 8-block model, near-view quality improves by \BlkDistNearGain{}\,dB over
the single-block baseline (\BlkDistEightNear{} vs.\ \BlkDistOneNear{}\,dB masked PSNR), while
far-view quality improves by \BlkDistFarGain{}\,dB (\BlkDistEightFar{} vs.\ \BlkDistOneFar{}\,dB),
consistent with block training providing additional capacity for
spatially localized internal structure.

\subsection{Generalization}
\label{sec:exp:generalization}

A practical compression pipeline must generalize across data scales,
spatial regions, and simulation types without per-dataset tuning.
All experiments in this section use the identical default pipeline
and hyperparameters unless stated otherwise.

\begin{figure}[t]
  \centering
  \includegraphics[width=\columnwidth]{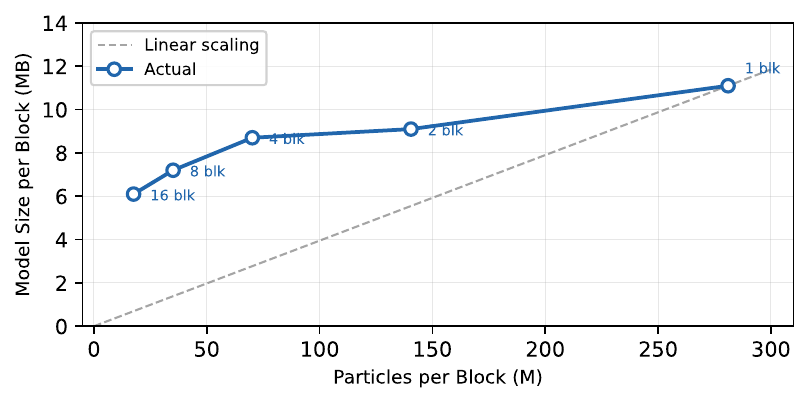}
  \caption{Per-block model size vs.\ particles per block.
  The dashed line shows perfect linear scaling (anchored at the single-block model).
  Actual model sizes grow sub-linearly: each block retains a
  base capacity regardless of particle count, so arbitrarily
  fine partitioning yields diminishing returns in total model size.}
  \label{fig:scale}
\end{figure}

Block training (\cref{sec:exp:block}) implicitly demonstrates scale
generalization: each block is trained independently, with per-block
particle counts spanning ${\sim}\ScaleSixteenParticlesM{}$M (16 blocks) to \HaccParticlesM{}M (single
block), a ${16\times}$ range. \cref{fig:scale} shows per-block Gaussian
counts scale sub-linearly with particles, from \ScaleSixteenGaussians{} (\ScaleSixteenParticlesM{}M)
to \ScaleOneGaussians{} (\HaccParticlesM{}M), so the pipeline automatically adapts
capacity to data size.

To verify that the method is not overfit to a particular spatial region,
we evaluate on four spatially disjoint HACC snapshots drawn from a
separate HACC dataset, each containing ${\sim}\HaccGenParticlesM{}$M
particles, comparable in size to our primary
\HaccParticlesM{}M-particle dataset.
\cref{tab:generalization} (top) shows remarkably consistent results:
mean PSNR is \HaccGenMeanPSNR{}\,dB with a standard deviation of only \HaccGenStdPSNR{}\,dB,
and all blocks achieve similar compression ratios ($\HaccGenCRLow$--$\HaccGenCRHigh\times$)
and model sizes ($\HaccGenSizeLow$--$\HaccGenSizeHigh$\,MB).

\begin{table}[t]
  \centering
  \caption{Generalization across spatial regions and simulation codes.
  \emph{Top:} four spatially disjoint HACC snapshots from a separate HACC dataset.
  \emph{Bottom:} dark-matter-only FIRE-2 cosmological box (\FireTwoParticlesM{}M particles, different simulation code).}
  \label{tab:generalization}
  \setlength{\tabcolsep}{4pt}
  \begin{tabular}{lcccc}
    \toprule
    \textbf{Dataset / Block} & \textbf{PSNR} & \textbf{Size} & \textbf{Gaussians} & \textbf{CR} \\
    \midrule
    HACC Block 0 & \HaccGenZeroPSNR{}\,dB & \HaccGenZeroSizeMB{}\,MB & \HaccGenZeroGaussians & $\HaccGenZeroCR\times$ \\
    HACC Block 1 & \HaccGenOnePSNR{}\,dB & \HaccGenOneSizeMB{}\,MB & \HaccGenOneGaussians & $\HaccGenOneCR\times$ \\
    HACC Block 2 & \HaccGenTwoPSNR{}\,dB & \HaccGenTwoSizeMB{}\,MB & \HaccGenTwoGaussians & $\HaccGenTwoCR\times$ \\
    HACC Block 3 & \HaccGenThreePSNR{}\,dB & \HaccGenThreeSizeMB{}\,MB & \HaccGenThreeGaussians & $\HaccGenThreeCR\times$ \\
    \midrule
    HACC Mean $\pm$ std & $\HaccGenMeanPSNR \pm \HaccGenStdPSNR$\,dB & \HaccGenMeanSizeMB{}\,MB & \HaccGenMeanGaussians & $\HaccGenMeanCR\times$ \\
    \midrule
    FIRE-2 & \FireTwoPSNR{}\,dB & \FireTwoSizeMB{}\,MB & \FireTwoGaussians & $\FireTwoCR\times$ \\
    \bottomrule
  \end{tabular}
\end{table}

We further evaluate on the dark-matter-only cosmological box from the
FIRE-2 public data release~\cite{wetzel2023public}, which contains
\FireTwoParticlesM{}M particles in a $\FireTwoDomain{}^3$ domain and is
produced by a different simulation code than HACC, with a different
domain size and clustering morphology.
The only adaptation is scaling ParaView's particle radius and
its sampling range proportionally to the domain size.
As shown in \cref{tab:generalization} (bottom), the pipeline achieves
\FireTwoPSNR{}\,dB PSNR with a compact \FireTwoSizeMB{}\,MB model ($\FireTwoCR\times$),
using fewer Gaussians (\FireTwoGaussians{} vs.\ ${\sim}$\HaccGenMeanGaussians{} for HACC),
consistent with the more concentrated matter distribution.

\subsection{Generalization to Unseen Views and Parameters}
\label{sec:exp:gen_view}

The results above generalize across \emph{datasets}; we further test
generalization across \emph{views and parameters} held out from training.
We use the single-block model (\GenModelGaussians{} Gaussians), whose
\GenReproPSNR{}\,dB mean full-image PSNR over the trained orbits
reproduces the single-block operating point.
\cref{tab:gen_view} reports masked PSNR at orbit radii never seen during
training, both interpolated between the \OrbitFar/\OrbitMid/\OrbitNear
training orbits and extrapolated beyond them, with per-frame $(r,\alpha)$
held in the trained range.
Unseen orbits show no generalization gap: masked PSNR stays within
\GenPoseSpreadDB{}\,dB of the trained orbits, and each interpolated
value falls between those of its neighboring trained orbits.
The one exception is aggressive near-extrapolation: at $0.35\times$,
closer than any training orbit, masked PSNR drops by \GenNearDropDB{}\,dB.
SSIM varies with orbit distance itself, because the foreground fraction
grows as the camera approaches, but shows the same pattern.

We then fix the orbit and push the radius and opacity factors outside
their trained range $[0.25, 1.75]$.
Quality degrades smoothly, with no discontinuity at the range boundary:
masked PSNR is \GenRadTwoPSNR{}\,dB at radius factor $2.0\times$ and
\GenRadTwoHalfPSNR{}\,dB at $2.5\times$, and \GenOpaTwoPSNR{}\,dB and
\GenOpaTwoFourPSNR{}\,dB at opacity factors $2.0\times$ and $2.4\times$,
continuing the in-range trend.
VizMapper extrapolates rather than collapses outside its training
envelope.

\begin{table}[t]
  \centering
  \caption{Generalization to unseen camera orbits (single-block model, masked PSNR
  and SSIM, $(r,\alpha)$ in trained range). Interpolated (\textit{int}) and far-extrapolated
  (\textit{ext}) orbits match the trained orbits (\textit{tr}); only aggressive
  near-extrapolation costs \GenNearDropDB{}\,dB.}
  \label{tab:gen_view}
  \setlength{\tabcolsep}{4pt}
  \begin{tabular}{lccccccc}
    \toprule
    Orbit & $1.3\times$ & $1.0\times$ & $0.85\times$ & $0.7\times$ & $0.6\times$ & $0.5\times$ & $0.35\times$ \\
          & \tiny ext & \tiny tr & \tiny int & \tiny tr & \tiny int & \tiny tr & \tiny ext \\
    \midrule
    Masked PSNR & \GenOrbitFarPSNR & \GenOrbitTrainFarPSNR & \GenOrbitInterpHiPSNR
      & \GenOrbitTrainMidPSNR & \GenOrbitInterpLoPSNR & \GenOrbitTrainNearPSNR
      & \GenOrbitNearPSNR \\
    SSIM & \GenOrbitFarSSIM & \GenOrbitTrainFarSSIM & \GenOrbitInterpHiSSIM
      & \GenOrbitTrainMidSSIM & \GenOrbitInterpLoSSIM & \GenOrbitTrainNearSSIM
      & \GenOrbitNearSSIM \\
    \bottomrule
  \end{tabular}
\end{table}

\begin{figure}[!t]
  \centering
  \includegraphics[width=\columnwidth]{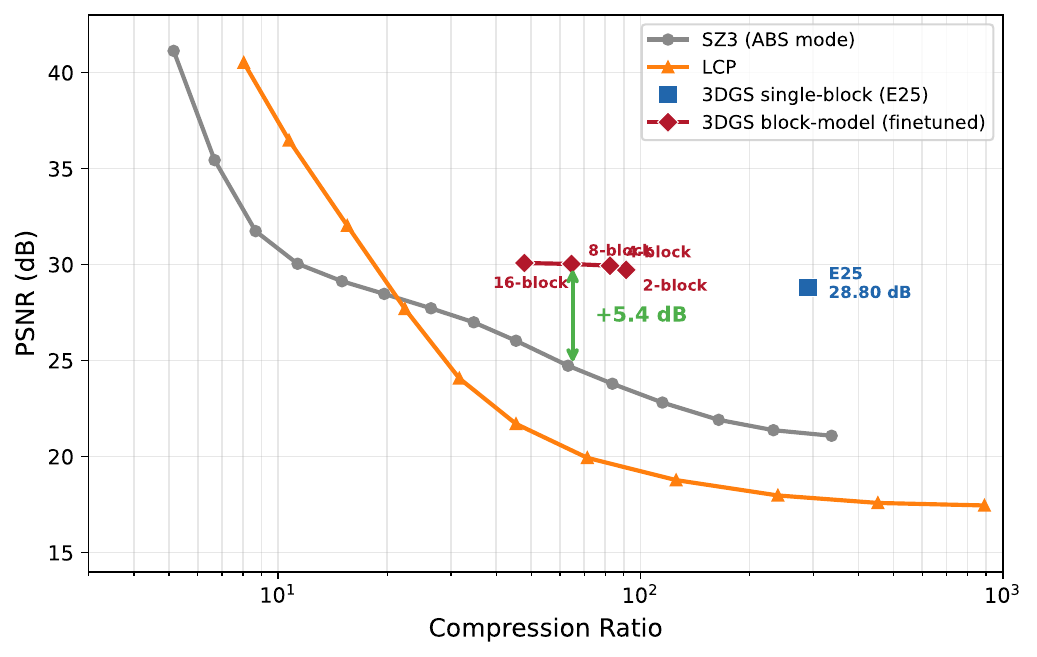}
  \caption{Rate-distortion comparison between our 3DGS models, SZ3, and LCP.
  Our models consistently outperform both baselines across the full
  compression range, achieving $\GainSzEb$\,dB over SZ3 at the 8-block
  operating point ($\EbCR\times$ CR) and $\GainSzSb$\,dB at the
  single-block point ($\SbCR\times$ CR).}
  \label{fig:rd}
\end{figure}

\begin{figure*}[!t]
  \centering
  \includegraphics[width=\textwidth]{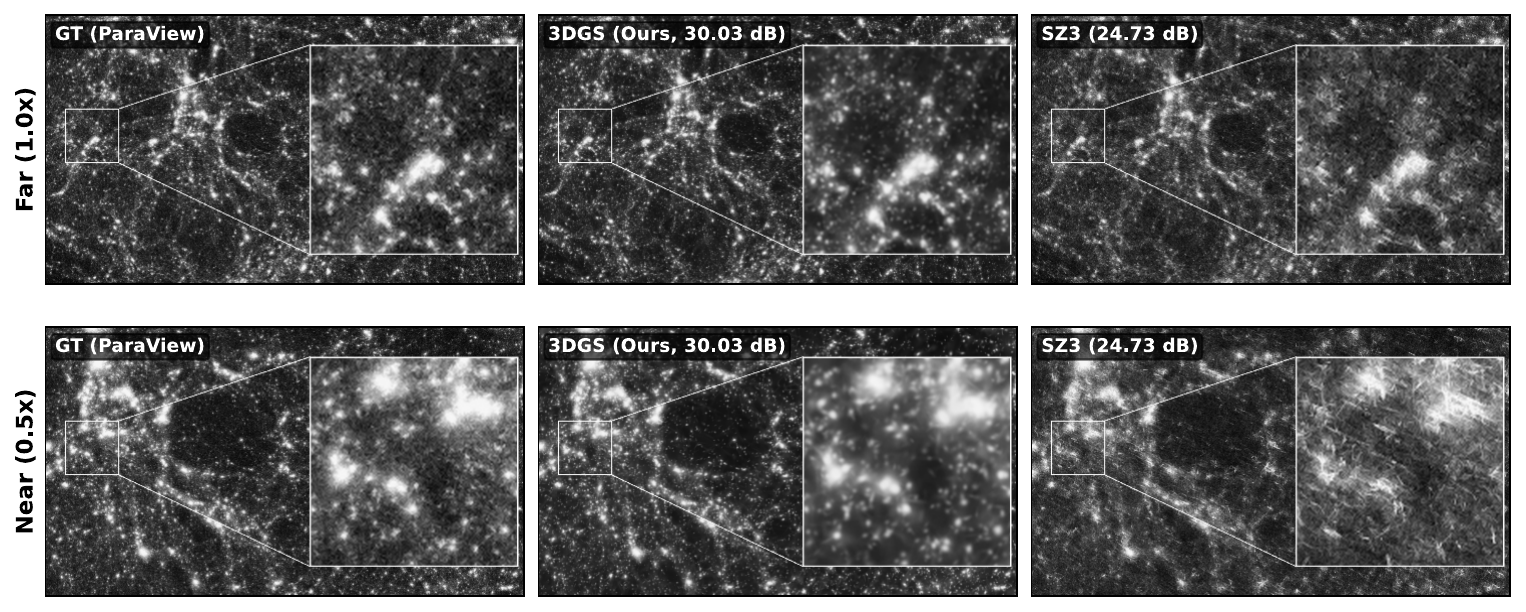}
  \caption{Visual comparison at far (top, orbit $\OrbitFar\times$) and near (bottom,
  orbit $\OrbitNear\times$) viewing distances, both at ${\sim}\EbCR\times$ compression.
  Left: ParaView ground truth.
  Center: our 8-block 3DGS model (\EbPSNR{}\,dB).
  Right: SZ3 at matched ratio (\QualSzPSNR{}\,dB), exhibiting axis-aligned
  streaking artifacts visible in the inset.
  These streaks arise from SZ3's per-axis 1D compression: $x$, $y$, $z$
  arrays are stored separately with different spatial continuity, so the
  predictor produces independent quantization errors along each axis
  that align as stripes. Images show the center 30\% crop for detail.}
  \label{fig:qualitative}
\end{figure*}

\subsection{Rate-Distortion: 3DGS vs.\ Baselines}
\label{sec:exp:rd}

\begin{table}[!t]
  \centering
  \caption{Rate-distortion comparison.
  Compression ratio (CR) is computed against raw position data (\HaccRawGB{}\,GB).
  PSNR is averaged over far/mid/near evaluation orbits.
  Both 3DGS variants outperform SZ3 and LCP by \GainSzLow{}--\GainSzHigh{}\,dB at matched ratios.}
  \label{tab:rd}
  \begin{tabular}{lccc}
    \toprule
    \textbf{Method} & \textbf{CR} & \textbf{Size} & \textbf{PSNR} \\
    \midrule
    SZ3 (matched to single) & $\sim$$\SbCR\times$ & $\sim$\SzSizeAtSbCR{}\,MB & $\sim$\SzPsnrAtSbCR{}\,dB \\
    SZ3 (matched to 8-blk) & $\sim$$\EbCR\times$ & $\sim$\SzSizeAtEbCR{}\,MB & $\sim$\SzPsnrAtEbCR{}\,dB \\
    LCP (matched to single) & $\sim$$\SbCR\times$ & $\sim$\SzSizeAtSbCR{}\,MB & $\sim$\LcpPsnrAtSbCR{}\,dB \\
    LCP (matched to 8-blk) & $\sim$$\EbCR\times$ & $\sim$\SzSizeAtEbCR{}\,MB & $\sim$\LcpPsnrAtEbCR{}\,dB \\
    \midrule
    3DGS single (ours) & $\SbCR\times$ & \SbSizeMB{}\,MB & \textbf{\SbPSNR}\,dB \\
    3DGS 8-blk + FT (ours) & $\EbCR\times$ & \EbSizeMB{}\,MB & \textbf{\EbPSNR}\,dB \\
    \bottomrule
  \end{tabular}
\end{table}

Our 3DGS models outperform both baselines by \GainSzLow{}--\GainSzHigh{}\,dB across the
full compression range (\cref{fig:rd,tab:rd}).
Our single-block model (${\sim}$\SbGaussians{} Gaussians) achieves \SbPSNR{}\,dB
PSNR at a compression ratio of ${\SbCR\times}$
(\HaccRawGB{}\,GB $\to$ \SbSizeMB{}\,MB).
At a comparable compression ratio, SZ3 achieves only
${\sim}\SzPsnrAtSbCR$\,dB, a gap of $\GainSzSb$\,dB.
Our 8-block model reaches \EbPSNR{}\,dB at ${\EbCR\times}$ compression
(\EbSizeMB{}\,MB), outperforming SZ3 by $\GainSzEb$\,dB at matched
compression ratios. LCP, a particle-specific compressor, reaches \LcpPsnrAtEbCR{}\,dB at $\EbCR\times$
and \LcpPsnrAtSbCR{}\,dB at $\SbCR\times$: gaps of $\GainLcpEb$\,dB and $\GainLcpSb$\,dB against our method.

Visual quality differences are most apparent at close range
(\cref{fig:qualitative}).
At ${\sim}\EbCR\times$ compression, our 3DGS model preserves the
large-scale cosmic web and local density variations, while SZ3
exhibits visible artifacts at close range where per-coordinate
quantization errors become apparent.

\subsection{Performance}
\label{sec:exp:perf}

\begin{table}[t]
  \centering
  \caption{Performance summary (RTX PRO 6000 Blackwell).
  ParaView renders \HaccParticlesM{}M particles; 3DGS uses \EbGaussians{} Gaussians (8-block model).
  Viz-adaptation cost: ParaView = combined $-$ camera-only (same session);
  3DGS = per-component timing with CUDA synchronization.
  3DGS components and combined are measured in independent benchmark
  passes, so VM\,+\,camera need not exactly equal combined.
  VizMapper replaces ParaView's full pipeline rebuild on each parameter
  change, accounting for the bulk of the ${\SpeedupOneK}\times$ end-to-end speedup.}
  \label{tab:perf}
  \setlength{\tabcolsep}{4pt}
  \resizebox{\columnwidth}{!}{%
  \begin{tabular}{lrrr}
    \toprule
    & \textbf{ParaView} & \textbf{3DGS (Ours)} & \textbf{Speedup} \\
    \midrule
    \multicolumn{4}{l}{\textit{Per-frame latency (1080p)}} \\
    \quad Viz-param adapt. & \PerfPvVizMs{}\,ms & \PerfGsVmMs{}\,ms & \textcolor{green!50!black}{$\SpeedupVizOneK\times$} \\
    \quad Camera render    &   \PerfPvCamMs{}\,ms   & \PerfGsRastMs{}\,ms & \textcolor{green!50!black}{$\SpeedupCamOneK\times$} \\
    \quad Combined         & \PerfPvCombMs{}\,ms & \PerfGsCombMs{}\,ms & \textcolor{green!50!black}{$\SpeedupOneK\times$} \\
    \midrule
    \multicolumn{4}{l}{\textit{Per-frame latency (4K)}} \\
    \quad Viz-param adapt. & \PerfPvVizFourKMs{}\,ms & \PerfGsVmFourKMs{}\,ms & \textcolor{green!50!black}{$\SpeedupVizFourK\times$} \\
    \quad Camera render    &   \PerfPvCamFourKMs{}\,ms   & \PerfGsRastFourKMs{}\,ms & \textcolor{green!50!black}{$\SpeedupCamFourK\times$} \\
    \quad Combined         & \PerfPvCombFourKMs{}\,ms & \PerfGsCombFourKMs{}\,ms & \textcolor{green!50!black}{$\SpeedupFourK\times$} \\
    \midrule
    \multicolumn{4}{l}{\textit{GPU memory}} \\
    \quad Inference (peak)        & & \InferMemMB{}\,MB  & \\
    \quad Training (peak, 6K)     & & \TrainPeakGB{}\,GB  & \\
    \midrule
    \multicolumn{4}{l}{\textit{Pipeline cost}} \\
    \quad Data loading      & \PvLoadSec{}\,s & \GsLoadMs{}\,ms  & \textcolor{green!50!black}{$\LoadSpeedup\times$} \\
    \quad Data prep (one-time, raw$\to$VTP) & & $\PrepVtpMin$\,min & \\
    \quad Training (single-block, incl.\ GT gen) & & \TrainTimeMin{}\,min & \\
    \quad Merge (8 blocks)                  & & \MergeTimeSec{}\,s & \\
    \quad Finetune (\FtIters{} iter)               & & \FtTimeMin{}\,min & \\
    \bottomrule
  \end{tabular}}
\end{table}

Our pipeline renders ${\SpeedupOneK\times}$ faster than ParaView at
1080p (\GsFpsOneK{}\,FPS vs.\ \PvFps{}\,FPS) with \InferMemMB{}\,MB GPU memory, and
trains end-to-end in \TrainTimeMin{} minutes on a single GPU (\cref{tab:perf}).
Both ParaView and our 3DGS pipeline perform two operations per frame:
adapting to visualization parameter changes (particle radius, opacity)
and rendering from the current camera.
\cref{tab:perf} decomposes per-frame cost into these two stages,
measured as pure rendering throughput (no disk I/O).
In ParaView, changing visualization parameters via the standard Python API
triggers a full pipeline rebuild over \HaccParticlesM{} million particles
(\PerfPvVizMs{}\,ms), which dominates frame time; camera-only rendering
costs \PerfPvCamMs{}\,ms and is resolution-independent, as vertex
processing for \HaccParticlesM{}M particles far exceeds the fragment
workload at either resolution.
In our pipeline, the VizMapper MLP replaces this rebuild with a \PerfGsVmMs{}\,ms
forward pass over \EbGaussians{} Gaussians, while the CUDA rasterizer renders
in \PerfGsRastMs{}\,ms at 1080p (\PerfGsRastFourKMs{}\,ms at $4$K).
The overall ${\SpeedupOneK\times}$ speedup at 1080p enables interactive
exploration at \GsFpsOneK{}\,FPS (\GsFpsFourK{}\,FPS at $4$K), with the 8-block
model using only \InferMemMB{}\,MB of GPU memory at inference.
A single-block model trains in \TrainTimeMin{} minutes; the 8-block variant
adds a \MergeTimeSec{}\,s merge and a \FtTimeMin{}-min finetune over eight parallel
runs. The per-block trainings are fully independent, so wall-clock stays
roughly constant given one GPU per block, while aggregate GPU-hours grow
linearly with block count (our two-GPU workstation ran them in four
batches, a hardware rather than method limit). Data loading drops $\sim$1000$\times$ (\PvLoadSec{}\,s ParaView VTP
$\to$ \GsLoadMs{}\,ms 3DGS PLY).
We account for this training cost transparently: the \TrainTimeMin{}\,min
\emph{already includes} ParaView ground-truth image generation for all stages,
and the only additional offline step is a one-time $\PrepVtpMin$\,min
raw$\to$VTP conversion that is shared across every camera configuration and every
baseline. This is a larger one-time cost than a single pass of a data-space
compressor (SZ3 compresses the \HaccParticlesM{}M-particle positions in
\SzCompressSec{}\,s with a \SzDecompressSec{}\,s decompress, and LCP in \LcpCompressSec{}\,s
with \LcpDecompressSec{}\,s, at $\varepsilon{=}0.08$), but ParticleGS amortizes it over
\emph{unlimited} interactive exploration: each subsequent camera or $(r,\alpha)$ change
costs a \PerfGsVmMs{}\,ms VizMapper pass plus rasterization, whereas every such change in
the data-space pipeline re-executes ParaView's \PerfPvVizMs{}\,ms rebuild over all
\HaccParticlesM{}M particles.

\subsection{Approximate Particle Recovery via GMM Sampling}
\label{sec:exp:recovery}

The learned Gaussians can be interpreted as a Gaussian Mixture Model
(GMM) from which particle positions can be sampled.
Large-scale spatial statistics are well preserved (density correlation
${>}\RecCorrThreshold$, $P(k)$ and $\xi(r)$ within ${\pm}10\%$), but individual
particle positions are not faithfully encoded (NN distance at best
\RecNNPctOfGT{}\% of GT).
This analysis tests whether the 3DGS representation preserves spatial
statistics of the original particle distribution, a stronger
requirement than visual fidelity.
\cref{tab:recovery_methods} compares four sampling strategies on
\HaccParticlesM{}M recovered particles from the 8-block model.
Opacity-weighted sampling (V0) achieves the highest density
correlation (\RecVZeroCorr), while volume-only weighting (V1, \RecVOneCorr) and
density-field guided sampling (V6, \RecVSixCorr) fail to recover meaningful
structure: opacity learned for rendering also encodes local particle
density.

\begin{figure}[!t]
  \centering
  \includegraphics[width=\columnwidth]{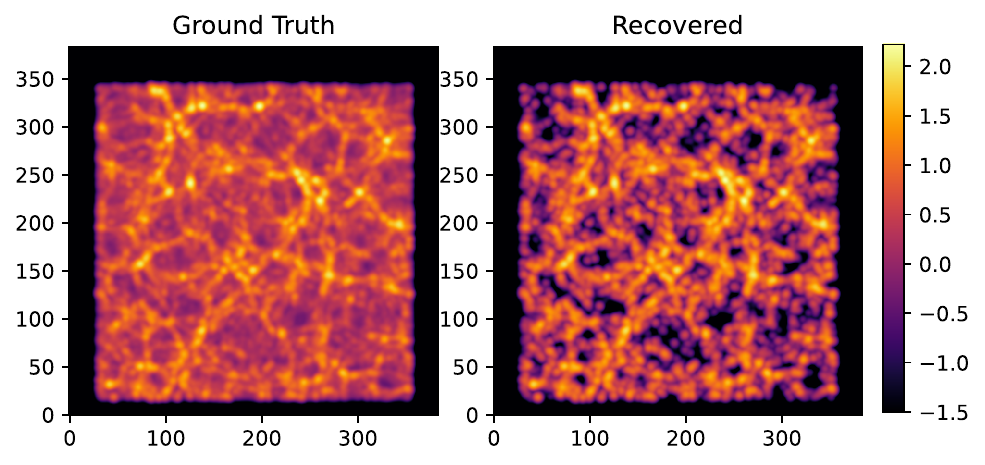}
  \caption{Density field central slice ($\log_{10}\!\rho/\bar{\rho}$)
  from the learned 8-block Gaussian mixture.
  Left: ground truth (\HaccParticlesM{}M particles). Right: recovered (\HaccParticlesM{}M particles).
  Large-scale structure (filaments, nodes, and voids) is clearly
  reproduced.}
  \label{fig:recovery:slices}
\end{figure}

The recovered particles are appropriate for \emph{density-domain}
analysis: density-field visualization and slicing
(\cref{fig:recovery:slices}), the matter power spectrum $P(k)$ (within
${\pm}10\%$ up to $k{=}\RecPkTenPctK$\,h/Mpc), the two-point correlation
$\xi(r)$ at large scales (within ${\pm}10\%$ for $r{>}\RecXiTenPctR$\,Mpc/h),
and identification of large-scale structure (filaments, nodes, voids).
They are \emph{not} appropriate for analyses that depend on individual
particles: nearest-neighbor and small-scale clustering statistics
(NN distance at best \RecNNPctOfGT{}\% of GT, and $\xi$ overestimated at
small $r$), per-particle tracking or identifiers, velocities and other
attributes, and small-scale halo finding.
This delineation follows directly from the representation being optimized
for visual rather than data-space fidelity, and we present recovery as an
approximate byproduct rather than a general-purpose decompression.

\begin{table}[!t]
  \centering
  \caption{Recovery algorithm comparison (\HaccParticlesM{}M particles, 8-block model).
  $\xi$ Dev: mean $|\xi_\text{rec}/\xi_\text{GT} - 1|$ over bins where $\xi_\text{GT} > 0$.}
  \label{tab:recovery_methods}
  \setlength{\tabcolsep}{4pt}
  \begin{tabular}{lcccc}
    \toprule
    \textbf{Method} & \textbf{Dens.\ Corr.} & \textbf{Dens.\ MAE}
    & \textbf{NN Mean} & \textbf{$\xi$ Dev} \\
    \midrule
    Baseline & \textbf{\RecVZeroCorr} & \RecVZeroMAE & \RecVZeroNNMean & \RecVZeroXiDev{}\% \\
    Volume-weighted & \RecVOneCorr & \RecVOneMAE & \RecVOneNNMean & \textbf{\RecVOneXiDev}\% \\
    Uniform ellipsoid & \RecVFourCorr & \textbf{\RecVFourMAE} & \textbf{\RecVFourNNMean} & \RecVFourXiDev{}\% \\
    Density field & \RecVSixCorr & \RecVSixMAE & \RecVSixNNMean & \RecVSixXiDev{}\% \\
    \bottomrule
  \end{tabular}
\end{table}

\begin{figure}[!t]
  \centering
  \includegraphics[width=\columnwidth]{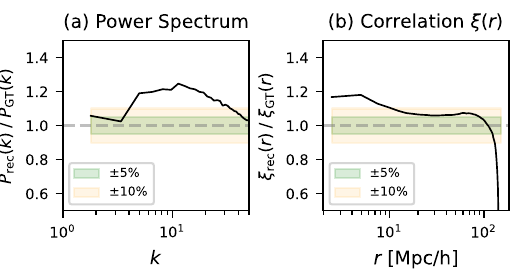}
  \caption{Cosmological spatial statistics~\cite{peebles2020large} of
  recovered particles (8-block model, \HaccParticlesM{}M particles).
  (a)~$P(k)$ remains within ${\pm}10\%$ up to $k = \RecPkTenPctK$\,h/Mpc.
  (b)~$\xi(r)$ agrees within ${\pm}10\%$ for $r > \RecXiTenPctR$\,Mpc/h;
  overestimates at small~$r$. Panels show ratios with ${\pm}5\%$ and ${\pm}10\%$ bands.}
  \label{fig:recovery:stats}
\end{figure}

\begin{table}[!t]
  \centering
  \caption{Particle recovery vs.\ block count (\HaccParticlesM{}M particles, V0 method).
  GT nearest-neighbor mean distance is \RecGtNNMean.}
  \label{tab:recovery_blocks}
  \begin{tabular}{rccccc}
    \toprule
    \textbf{Blocks} & \textbf{Dens.\ Corr.} & \textbf{MAE} & \textbf{NN Mean} & \textbf{$\xi$ Dev} \\
    \midrule
    1  & \RecBlkOneCorr & \RecBlkOneMAE & \RecBlkOneNNMean & \RecBlkOneXiDev{}\% \\
    2  & \RecBlkTwoCorr & \RecBlkTwoMAE & \RecBlkTwoNNMean & \RecBlkTwoXiDev{}\% \\
    4  & \RecBlkFourCorr & \RecBlkFourMAE & \RecBlkFourNNMean & \RecBlkFourXiDev{}\% \\
    8  & \textbf{\RecBlkEightCorr} & \RecBlkEightMAE & \RecBlkEightNNMean & \RecBlkEightXiDev{}\% \\
    16 & \textbf{\RecBlkSixteenCorr} & \textbf{\RecBlkSixteenMAE} & \textbf{\RecBlkSixteenNNMean} & \textbf{\RecBlkSixteenXiDev}\% \\
    \bottomrule
  \end{tabular}
\end{table}

\begin{figure}[!t]
  \centering
  \includegraphics[width=\columnwidth]{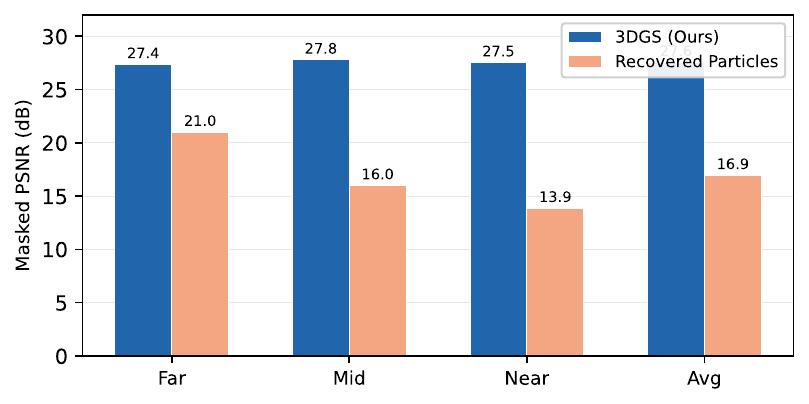}
  \caption{Three-way comparison: 3DGS rendering vs.\ recovered-particle
  rendering (ParaView), both evaluated against ground truth (masked PSNR).
  3DGS rendering consistently outperforms recovered-particle rendering
  by ${\sim}\ThreeWayGap$\,dB, confirming that the learned Gaussian
  representation captures visual structure beyond what the recovered
  particles preserve.}
  \label{fig:three_way}
\end{figure}

Large-scale density structure is well reproduced
(\cref{fig:recovery:slices}).
We deposit both ground-truth and recovered particles onto a $\RecGridSize^3$
grid to compute the density contrast field
$\delta = \rho/\bar{\rho} - 1$; large-scale
structure (filaments, nodes, and voids) is clearly reproduced.
The density PDF tracks the ground truth at
overdensities ($\delta > 0$) but diverges in voids ($\delta < -1$),
where the Gaussian mixture under-samples sparse regions.
The power spectrum $P(k)$, the primary summary statistic in
cosmological clustering analyses, remains within ${\pm}10\%$ of
the ground truth up to $k = \RecPkTenPctK$ for the 8-block model
(\cref{fig:recovery:stats}a).
The two-point correlation function $\xi(r)$, the real-space
counterpart of $P(k)$, agrees within ${\pm}10\%$ at separations
$r > \RecXiTenPctR$ (\cref{fig:recovery:stats}b), but overestimates
correlations at small~$r$ (up to $+18\%$ at $r \approx 5$),
consistent with clustering around Gaussian centers.

Increasing the block count improves recovery quality
(\cref{tab:recovery_blocks}): density correlation rises from \RecBlkOneCorr{}
(1~block) to \RecBlkEightCorr{} (8~blocks), as more Gaussians provide finer
spatial resolution for the mixture.
However, the nearest-neighbor (NN) mean distance remains well
below the ground truth (best: \RecBlkSixteenNNMean{} vs.\ GT~\RecGtNNMean{} for 16~blocks),
indicating that recovered particles cluster near Gaussian centers
rather than reproducing true inter-particle spacing.
Increasing the sampling spread ($1$--$3\times$ scale factor)
improves NN distance toward the GT value without substantially
affecting density correlation ($\Delta < 0.01$).

Rendering recovered particles through ParaView produces far worse
images than direct 3DGS rendering (\cref{fig:three_way}):
3DGS (\ThreeWayGsAvg{}\,dB) outperforms recovered-particle renderings
(\ThreeWayRecAvg{}\,dB) by over \ThreeWayGap{}\,dB, confirming that the learned
representation is optimized for appearance, not particle positions.

\section{Conclusion}
\label{sec:conclusion}

We have presented ParticleGS, a visualization-aware 3D Gaussian Splatting
framework for compressing and rendering large-scale scientific particle data.
Our approach learns a compact scene representation directly optimized for
rendered image quality, rather than point-wise data accuracy.
The VizMapper network enables a single model to adapt to different
visualization parameters without retraining, and spatial block
training scales model capacity to large datasets.
The complete pipeline trains in \TrainTimeMin{} minutes on one GPU.

ParticleGS has limitations at two levels.
At the implementation level, approximating ParaView's point-sprite
rendering with Gaussian alpha compositing imposes a quality ceiling,
and VizMapper inference is the per-frame throughput bottleneck.
Three deeper limitations bound the method's scope.
First, fidelity is highest within the trained viewing envelope: the
camera orbits and radius/opacity ranges seen during training are
reproduced faithfully, while quality degrades smoothly but measurably
outside this envelope (\cref{sec:exp:gen_view}), so arbitrary
interior fly-throughs are not fully covered.
Second, block training with global fine-tuning provides a scaling path:
blocks train in parallel at constant per-block capacity, and the
fine-tuning stage corrects inter-block boundaries.
We have demonstrated this path up to \HaccParticlesM{}M particles, but
not yet at the trillion-particle, multi-terabyte snapshots of the
largest campaigns.
Third, particle recovery is approximate and density-domain only
(\cref{sec:exp:recovery}): it supports large-scale structural and
statistical analysis, not per-particle positions, attributes, or
identifiers.

These limitations also delineate where the representation is
trustworthy.
It is reliable for qualitative exploration: inspecting large-scale
morphology, locating filaments, nodes, and voids, comparing density
fields across regions, and navigating camera and radius/opacity within
the trained ranges.
In these tasks the quantity of interest is the visual density field
that training directly optimizes.
It can be \emph{misleading} for quantities the rendered images do not
constrain: per-particle positions or identifiers, fine-scale clustering
and nearest-neighbor statistics, sharp features near block boundaries,
and extrapolation far outside the trained envelope, where the rendered
loss provides no quality guarantee.
Users should treat ParticleGS as an interactive visual surrogate, not
a source of ground-truth measurements.

We benchmark against ParaView as the representative general-purpose
visualization system, but the data-to-render bottleneck is not specific
to ParaView.
Parallel and in-situ particle renderers such as vl3 and Galaxy, the
VTK-m/Ascent stack, and level-of-detail point-cloud renderers in the
style of Potree all pay the same per-view pipeline cost over hundreds
of millions of particles, and would be informative additional points
of comparison.

Future work will extend the framework to time-varying simulations,
rendering modes beyond point-Gaussian density rendering, and additional
scientific domains.

The complete pipeline, experiment configurations, and the artifact
reproducing every reported number are available at
\url{https://github.com/BoJiang03/ParticleGS}.

\section*{Acknowledgment}
This material was supported by the U.S.~Department of Energy,
Office of Science, Advanced Scientific Computing Research (ASCR),
under contract DE-AC02-06CH11357, and by the National Science
Foundation under Grants OAC-2311875, OAC-2514036, and OAC-2513768.
This research used resources of the Argonne Leadership Computing
Facility, a U.S.~Department of Energy Office of Science user facility,
and the Bebop cluster operated by the Laboratory Computing Resource
Center at Argonne National Laboratory.
The authors used Claude (Anthropic) to assist with manuscript editing
and artifact-script development; all technical content, experiments,
and conclusions are the authors' own.

\newpage
\bibliographystyle{IEEEtran}
\bibliography{refs}

\end{document}